\documentclass[journal]{IEEEtran}
\usepackage{array,makecell}
\usepackage[caption=false,font=footnotesize]{subfig}
 \usepackage{dblfloatfix}
\usepackage{cite}
\usepackage{url}
\usepackage{amsmath,amssymb,amsfonts}
\usepackage{algorithmic}
\usepackage{graphicx}
\graphicspath{{C:/Users/bhat_po/Desktop/work_in_progress/RI_TCI/tci_RPS_second_revision/images/}}
\usepackage{textcomp}
\usepackage[table]{xcolor}
\usepackage{caption}
\usepackage{tabularx}
\usepackage{multirow}
\usepackage{xr}
\newcommand{\field}[1]{\mathbf{#1}}
\newcommand{\R}{\field{R}}

\newcommand{\norm}[1]{\Vert{#1} \Vert_2^2}

\newcommand{\sg}{subgradient }

\newcommand{\sgs}{subgradients }

\newcommand{\Ac}{A^C}
\newcommand{\Acs}{(A^C)^*}
\newcommand{\Ar}{A^R}
\newcommand{\Ars}{(A^R)^*}

\newcommand{\pc}{p^C}
\newcommand{\pr}{p^R}
\newcommand{\xc}{x^C}
\newcommand{\xr}{x^R}
\newcommand{\yc}{y^C}
\newcommand{\yr}{y^R}
\newcommand{\yb}{y^B}
\newcommand{\Bs}{B^*}
\usepackage{optidef}

\begin{document}
\bstctlcite{IEEEexample:BSTcontrol}
\title{Region-of-Interest Prioritised Sampling for Constrained Autonomous Exploration Systems}
\author{Protim Bhattacharjee, Martin Burger, Anko B{\"o}rner, and Veniamin I. Morgenshtern
        % <-this % stops a space
        \thanks{P. Bhattacharjee is with the Department of Optical Sensor Systems, German Aerospace Center (DLR), Berlin, Germany (e-mail: protim.bhattacharjee@dlr.de).}% <-this % stops a space
        \thanks{M. Burger is with the Department of Mathematics, Friedrich Alexander Universit{\"a}t Erlangen-N{\"u}rnberg, Erlangen, Germany (e-mail: martin.burger@fau.de).}%
\thanks{ A. B{\"o}rner is with the Department of Optical Sensor Systems, German Aerospace Center (DLR), Berlin, Germany (e-mail: anko.boerner@dlr.de).}%
\thanks{ V. I. Morgenshtern is with the Department of Electrical-Electronic-Communication Engineering, Friedrich Alexander Universit{\"a}t Erlangen-N{\"u}rnberg, Erlangen, Germany (e-mail: veniamin.morgenshtern@fau.de).}}


\markboth{Journal of \LaTeX\ Class Files,~Vol.~14, No.~8, August~2015}%
{Shell \MakeLowercase{\textit{et al.}}: Bare Demo of IEEEtran.cls for IEEE Journals}



\maketitle

\begin{abstract}
Goal oriented autonomous operation of space rovers has been known to increase scientific output of a mission. In this work we present an algorithm, called the RoI Prioritised Sampling (RPS), that prioritises Region-of-Interests (RoIs) in an exploration scenario in order to utilise the limited resources of the imaging instrument on the rover effectively. This prioritisation is based on an estimator that evaluates the change in information content at consecutive spatial scales of the RoIs without calculating the finer scale reconstruction. The estimator, called the Refinement Indicator (RI), is motivated and derived. Multi-scale acquisition approaches, based on classical and multi-level compressed sensing, with respect to the single pixel camera architecture are discussed. The performance of the algorithm is verified on remote sensing images and compared with the state-of-the-art multi-resolution reconstruction algorithms. At the considered sub-sampling rates the RPS is shown to better utilise the system resources for reconstructing the RoIs.

\end{abstract}


% \fi
\IEEEpeerreviewmaketitle

\section{Introduction}
\label{sec:introduction}
\IEEEPARstart{I}{mage} acquisition in resource constrained environments is a challenging task. Activities such as space exploration and investigation of disaster sites are carried out by robotic platforms that have limited electrical power and payload capabilities. Being far away from the ground-station/human operator, exploration platforms, e.g., rovers and copters, require autonomous operation for achieving mission objectives. This necessitates new on-board sensing protocols, object and event detection algorithms, and data processing frameworks~\cite{autonomous1, AEGIS}. The imaging systems on autonomous exploration platforms are required to acquire maximum information about the scene with minimum resources. Limitations on electrical power impact the number of measurements that can be acquired and the number of computations that can be performed by the imaging system. For such constrained systems efficient utilization of their limited resources is important. The main purpose of the exploring rovers is to provide a survey of the scene, so as to recognise areas of importance, and possibly, to provide initial data for further examination by more precise instruments. This reduces the requirement of acquiring the entire scene at the highest resolution of the camera on-board the robotic platform. Regions-of-Interest (RoIs) can be recognized and acquired at the resolution of the camera leaving the background at lower resolutions. 

%\IEEEpubidadjcol

To ensure that measurements are expended on RoIs with more information it is necessary to prioritise RoIs for acquisition on the basis of their information content. Moreover, this should be performed in an online manner. However, existing methods of acquisition and reconstruction in constrained systems either try to reconstruct the entire scene at low resolutions~\cite{stoneTransform, ampMR} or require pre-defined resolutions for each RoI~\cite{multiResCompSensing}. These methods do not dynamically distribute their measurements among RoIs based on their information content. Also, they may require prior information (human intervention or some other source) to decide on the spatial resolution of each RoI. 

Dynamic real-time multi-resolution RoI acquisition based on the information content of the RoI is an important step forward for introducing mission oriented autonomy to exploration systems (E4~level autonomy)~\cite{ergo}. There are various ways of characterising the information content of a scene or RoI. In this work an adaptive multi-scale approach is used. An estimator, called the Refinement Indicator (RI), is developed to estimate changes in the information content of a RoI at consecutive spatial resolutions. The RoI with the largest change is refined to a finer spatial resolution. With the help of the RI and the basic framework for sampling suggested in~\cite{MCS} for measurement-constrained systems an algorithm for multi-scale sampling is proposed and named RoI Prioritised Sampling (RPS). The RI provides a structured way to prioritise RoIs for acquisition at the cost of a small number of overhead measurements at each spatial resolution. These additional measurements are also used for reconstruction of the RoIs using compressed sensing~\cite{CRT1, donoho1}. 

The main contributions of this work are $1)$ the development of an online estimator for information change across spatial scales for RoIs, $2)$ the design of an acquisition algorithm for constrained systems that incorporates this estimator to prioritise RoIs for acquisition and accordingly distributes the limited measurement budget, and $3)$ the proposal of multi-scale acquisition methodologies based on classical~\cite{fourTimesSparsity, donohoExtension} and multi-level compressed sensing~\cite{donohoExtension, roman2014asymptotic, asymptoticSparsity}. The algorithm is tested on % real-world%
images from airborne Earth observation sensor platforms and the Mars Science Laboratory (MSL) on-board the Curiosity rover on Mars. The proposed algorithm is found to perform better at reconstructing informative RoIs than existing algorithms at the considered sub-sampling rates. 

The next section reviews existing work on autonomous space exploration and compressed sensing based acquisition in constrained systems. Sec.~\ref{sec:cameraModel} discusses the camera model used in this work and the sampling methodology suggested in~\cite{MCS}. The RI is motivated and derived in Sec.~\ref{sec:RI}. Sec.~\ref{sec:results} discusses the datasets and results as applied to these datasets. The article ends with a conclusion and the direction for future work.

\section{Literature Review}
\label{sec:review}
\subsection{Autonomous Space Exploration}

Autonomous Exploration for Gathering Increased Science (AEGIS)~\cite{AEGIS} was developed for the Opportunity rover~\cite{Oppurtunity} on Mars to increase the scientific output of the mission by allowing autonomous acquisition of mission relevant data. The aim of the AEGIS system is to detect rocks by analysing images from the navigation camera of the rover and to direct scientific instruments, like the ChemCam~\cite{ChemCam}, to acquire scientific data from the surroundings of the rover without consultation with Earth scientists. Earlier all images were downlinked and analysed on Earth and then the rovers were instructed to perform extra measurements or revisit an area from which they had moved away. However, with the AEGIS system new scientific information could be gathered autonomously by the rover leading to increased scientific throughput~\cite{AEGIS_Result}. The current work builds on the idea of autonomous scientific data collection and provides a method for prioritising (ranking) the RoIs based on spatial frequency variations without any prior knowledge about the surrounding terrain. 

\subsection{Compressed Sensing for Constrained Systems}
Multi-resolution techniques have previously been used to address concerns of constrained systems. In~\cite{stoneTransform} the aim is to provide low resolution previews in computationally-constrained and data-streaming systems where computational machinery for compressed sensing reconstruction is not available. The authors propose an orthogonal sampling matrix known as the Sum-To-One (STOne) transform. This matrix allows one to recover low resolution previews at the Nyquist-rate in real-time.  The sampling matrix contains only $\pm 1$ entries and is suitable for implementation on a~bi-stable Digital Micromirror Device (DMD) used in Single Pixel Cameras (SPCs). The nature of sampling is such that when computational resources are available (at the ground/base-station) the measurements used for the preview can also be used to generate a high resolution image at sub-Nyquist rates. The previews can be generated at any resolution that is a power of 2 limited by the size of the DMD. The STOne transform thus tries to acquire and reconstruct the entire scene in a computationally constrained system.  This scenario can also be thought of as a measurement-constrained problem where the limited measurements are designed such that the exploration system can produce low resolution images for autonomous functions and high resolution recovery is possible only at a ground-station. Authors in~\cite{ampMR} cast the measurement-constrained reconstruction problem into a Multi-Resolution Approximate Message Passing (MR-AMP) framework. The measurement model is transformed in such a way that the low resolution image is an optimization variable. Properly designed down-sampling and up-sampling matrices are required to map the image from the native resolution of the sensor to the lower resolution and vice-versa. A down-sampling factor can be used to choose the down-sampling ratio according to the available number of measurements. As in the case of the STOne transform, MR-AMP tries to reconstruct the complete scene at the desired resolution. Multi-resolution in RoIs in the context of compressed sensing camera architectures was proposed in~\cite{multiResCompSensing}. Measurements for the entire scene are acquired at the native resolution of the DMD. The original image is then split into a number of pre-defined RoIs, and with the help of down-sampling and up-sampling matrices, different RoIs are reconstructed at different resolutions. The RoIs need to be chosen off-line and are reconstructed from the same measurements. The three techniques described model the reconstruction problem in a manner that enables multi-resolution recovery. In~\cite{MCS}, the authors propose a different acquisition approach for measurement-constrained systems. A three-step methodology for acquiring RoIs in a scene is proposed that includes a ``Low resolution acquisition and reconstruction" step, ``RoI detection and segmentation" step, and a ``Multi-level sampling" step. The entire process is online; it is performed in-situ without any external guidance or human intervention. The segmentation procedure is performed on the basis of mission objectives and multi-level compressed sensing is used to recover the RoIs. Each RoI is assigned a measurement budget on the basis of its size and is acquired individually. However, the RoI selection procedure is empirical and the algorithm acquires RoIs based on their sizes and tries to reconstruct each RoI at the native resolution of the DMD in one shot with no regard to the information content of the RoI. We propose to estimate the change in information content across spatial scales in the RoIs in an online fashion and distribute the measurement budget accordingly. A step-by-step increase in resolution is obtained in the RoIs by acquiring them at different resolutions instead of reconstructing them directly from measurements at the native resolution of the DMD.

\subsection{Bregman Distance}
\begin{figure}[!t]
\centering
\includegraphics[width=\columnwidth, height=5cm, keepaspectratio]{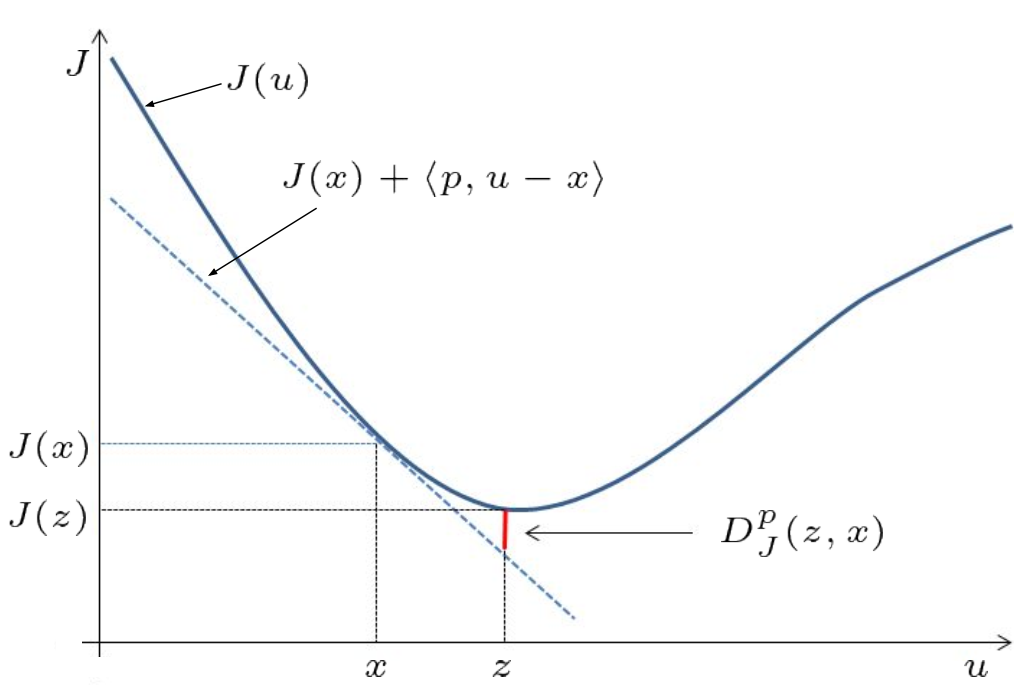}
\caption[Example Bregman distance.]{Bregman Distance is the distance, displayed in red, between $J(z)$ and the tangent to $J(\cdot)$ at $x$ evaluated at $z$. }
\label{reviewFig18}
\end{figure}

In this work, Bregman distance~\cite{burger2016} will be used to calculate error estimates. For a convex functional $J$ with a \sg $p \in \partial J(x)$, the (generalized) Bregman distance between two vectors $z$ and $x$ is defined as ${D_J^{p}(z,x) = J(z) - J(x) - \langle p, z-x\rangle,}$ where $\partial J(x)$ is the subdifferential of $J$ at $x$. The Bregman distance is the distance between $J(z)$ and the tangent to $J$ at $x$ evaluated at $z$. This is shown in Fig.~\ref{reviewFig18}. For a convex functional~$J$, the non-negativity is evident from the definition of $J$. However, the Bregman distance is not necessarily symmetric; a symmetric Bregman distance~\cite{burger2016} with respect to two \sgs of $J$, $p \in \partial J(z)$ and $q \in \partial J(x)$, is defined as ${ D^{p,q}_{J}(z,x) = D^{p}_{J}(x,z) + D^{q}_{J}(z,x) = \langle p - q, z - x\rangle}$. This symmetric form of the Bregman distance will be used to estimate the error in reconstruction of RoIs at various spatial scales. Similar methods for deriving error estimates for regularisation problems~\cite{burger2018} and image restoration problems~\cite{burger2007} based on Bregman distances have been studied previously. For $l_1$ regularised problem, common in compressed sensing, the Bregman distance is related to the sparsity of the signal~\cite[Proposition~8.2]{burger2007}.

\section{Single Pixel Camera Model and Acquisition Methodology}
\label{sec:cameraModel}

\subsection{Single Pixel Camera}
\label{sec:spc}

\begin{figure}[!t]
\centering
\includegraphics[width=\columnwidth, height = 5cm, keepaspectratio]{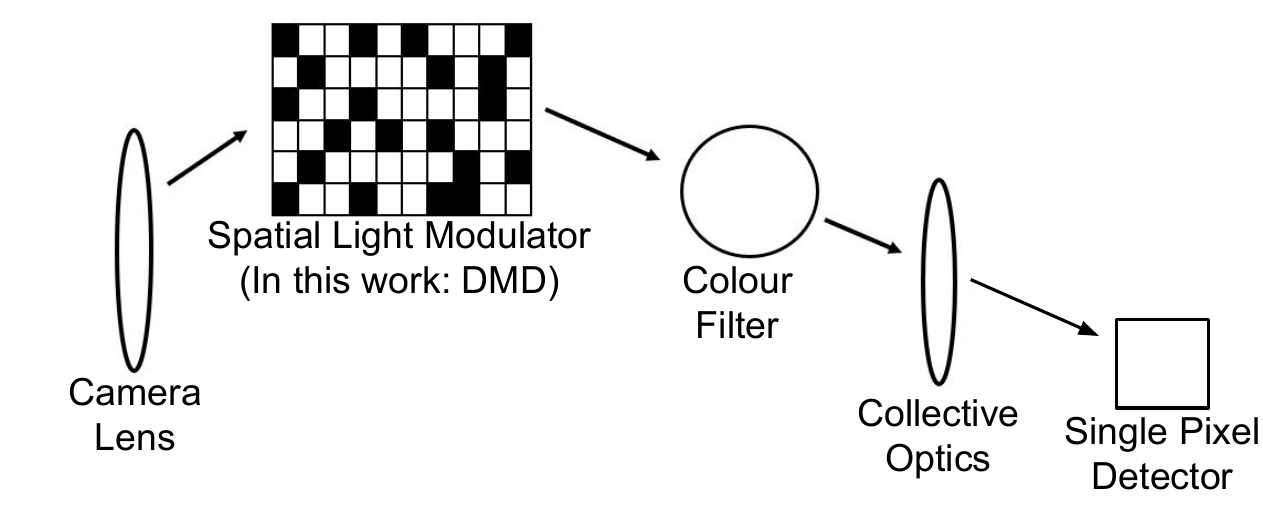} 
\caption[single Pixel Camera.]{Single Pixel Camera model.}
\label{fig:SPC}
\end{figure}

Different applications require acquisition of the scene at different wavelengths of the electromagnetic spectrum. For example, fire or temperature detection of objects require near-to-mid infrared imaging~\cite{firebird}. Terahertz radiation is used in security applications~\cite{terahertzSecurityUsecase}. Building large sensor arrays for imaging in longer wavelengths is difficult and expensive. The alternative is to use a single detector appropriate for the incident wavelength. Such a detector when calibrated is known as a radiometer~\cite{hp3Rad, MARA, terahertzSven}. A mechanical system can be used to raster scan the scene with the radiometer to generate a 2D image of the scene. However, in space applications mechanical systems are susceptible to damage due to stress and vibrations during rocket launch and landing. SPCs~\cite{spc_barniuk} are best suited for such exploration scenarios as they avoid large sensor arrays and mechanical scanning. In this work, the SPC architecture is used as the model for the imaging system. It consists of a camera lens that focusses the incoming radiation onto the spatial light modulator that is placed at the virtual imaging plane of the camera lens. The DMD~\cite{dmdTI2} is used as the spatial light modulator. It provides the necessary ``software" scan of the scene replacing mechanical raster scanning. The modulated radiation from the DMD may be filtered through a wavelength selective filter or a colour filter. The collective optics focusses the modulated filtered radiation onto the single pixel detector. At each measurement step, the DMD implements a measurement mask and the incoming radiation is encoded with the mask. This encoded radiation is received by the detector and generates one measurement. A sequence of such masks is displayed on the DMD with each mask generating a new measurement. A bi-stable DMD where the micromirrors can only have an `ON' or `OFF' state is used. By convention, the `ON' state directs the incident radiation towards the detector and the `OFF' state deflects radiation away from the detector. For a measurement mask consisting of $0/1$ entries, the ``$0$" pixels are mapped to the `OFF' state and the ``$1$" to the `ON' state. Such masks are used to perform random-macro-pixel sampling as discussed in the next subsection. To implement $\pm1$ entries of a measurement mask, two physical measurement cycles are required on the DMD. The first cycle maps all $+1$s to the `ON' state and the $-1$s to the `OFF' state. In the next measurement cycle all the $+1$s are mapped to the `OFF' state and the $-1$s to the `ON' state. Subtracting the two measurements thus obtained leads to one measurement generated by the measurement mask with $\pm1$ entries. Walsh transforms~\cite{orthogonalTrasnformsAhmed} require such implementations on the DMD. The basic single pixel camera model is shown in Fig.~\ref{fig:SPC}.

\begin{figure*}[!t]
\centering
\includegraphics[scale = 0.62, keepaspectratio]{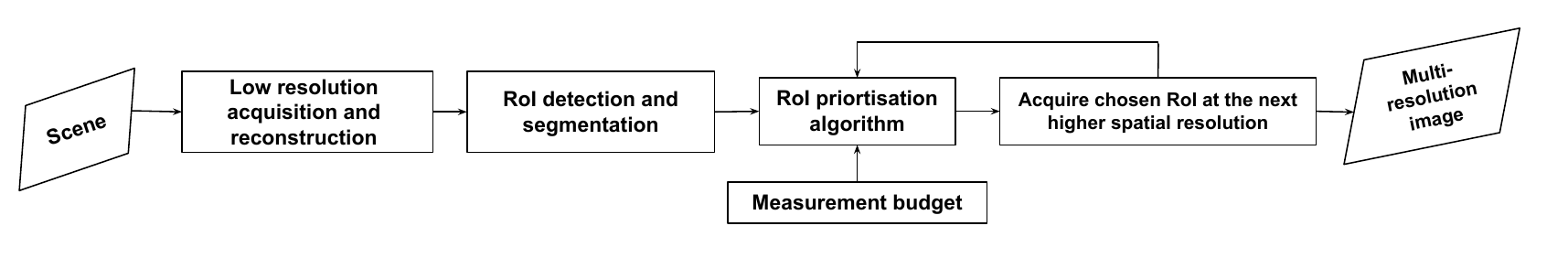} 
\caption[Proposed acquisition framework.]{Proposed acquisition methodology.}
\label{fig:basicAlgo}
\end{figure*}

The scene is acquired using the three step acquisition methodology proposed in~\cite{MCS}. The first step is the low resolution acquisition that is performed using macro pixels formed by binning micromirrors of the DMD. A $8 \times 8$ macro pixel is formed by binning 64 micromirrors underlying a $8 \times 8$ window on the DMD and assigning them the same `ON' or `OFF' state. A $256 \times 256$ DMD has 1024 non-overlapping $8 \times 8$ macro pixels. Similarly, it has 4096 non-overlapping $4 \times 4$ macro pixels. For each measurement, these macro pixels are randomly assigned the value of $0/1$ or $\pm1$ to simulate random-macro-pixel sampling. Details on the low resolution reconstruction will follow in Sec.~\ref{sec:csAlgos}. The second step is the detection and segmentation based on the low resolution reconstruction. This process depends on the mission objectives. For example, in infrared imaging, one may use temperature as a parameter for detection and segmentation of RoIs; in visible wavelength imaging, one may use the contrast of the scene. The process is same as in~\cite{MCS}. First, the low resolution reconstruction is resized to its macro pixel image dimensions. For example, a $256 \times 256$ image with a macro pixel size of $8\times8$ would be resized to a $32 \times 32$ image by averaging each $8 \times 8$ block in the low resolution reconstruction. A seed pixel is assigned to be the brightest pixel in the resized image. A square region is grown around the seed pixel by increasing the region size by one pixel in each direction until the contrast of the region is higher than a user defined threshold. This forms one RoI. A new seed pixel is assigned to be the brightest pixel outside the already selected region. Regions within a merge radius of each other are coalesced forming a larger RoI. This seed pixel selection and region growing is continued until the user defined total number of RoIs is reached. Once all the RoIs are selected, the segmented image is resized to the original size of the low resolution reconstruction, equal to the DMD size. Thereafter, each RoI is expanded to its immediate higher dyadic size. If this leads to more than 99\% overlap between RoIs, they are merged once more. The total number of regions to be selected, the contrast cut-off, the merge radius, and the overlap in the dyadic sizes are user defined. Fig.~\ref{segProc} in the supplementary material describes the process. The third step of the algorithm in~\cite{MCS} is the sequential acquisition of RoIs detected in the second step at the full resolution of the DMD using multi-level sampling~\cite{roman2014asymptotic}. The RoIs are arranged in a decreasing manner according to their size and are resolved to their full resolution in a one-shot manner. In this work we propose to estimate the change of information content in spatial scales of the RoIs and develop a procedure for step-by-step increase in the resolution of the RoIs. We propose two ways in which this step-by-step increase can be performed. One is through multi-scale random-macro-pixel sampling and the other is through the Walsh transform. The RoIs are acquired sequentially based on their information content and the number of measurements needed at each spatial scale depends on the RoI size and the current spatial resolution. The workflow of the proposed acquisition methodology is shown in Fig.~\ref{fig:basicAlgo}. We first discuss the methods of acquisition of higher spatial frequencies as it will be required for the development of the estimator.

\subsection{Multi-scale Random-macro-pixel  Sampling}

\begin{figure}[!t]
\centering
\includegraphics[width=\columnwidth, height=5cm, keepaspectratio]{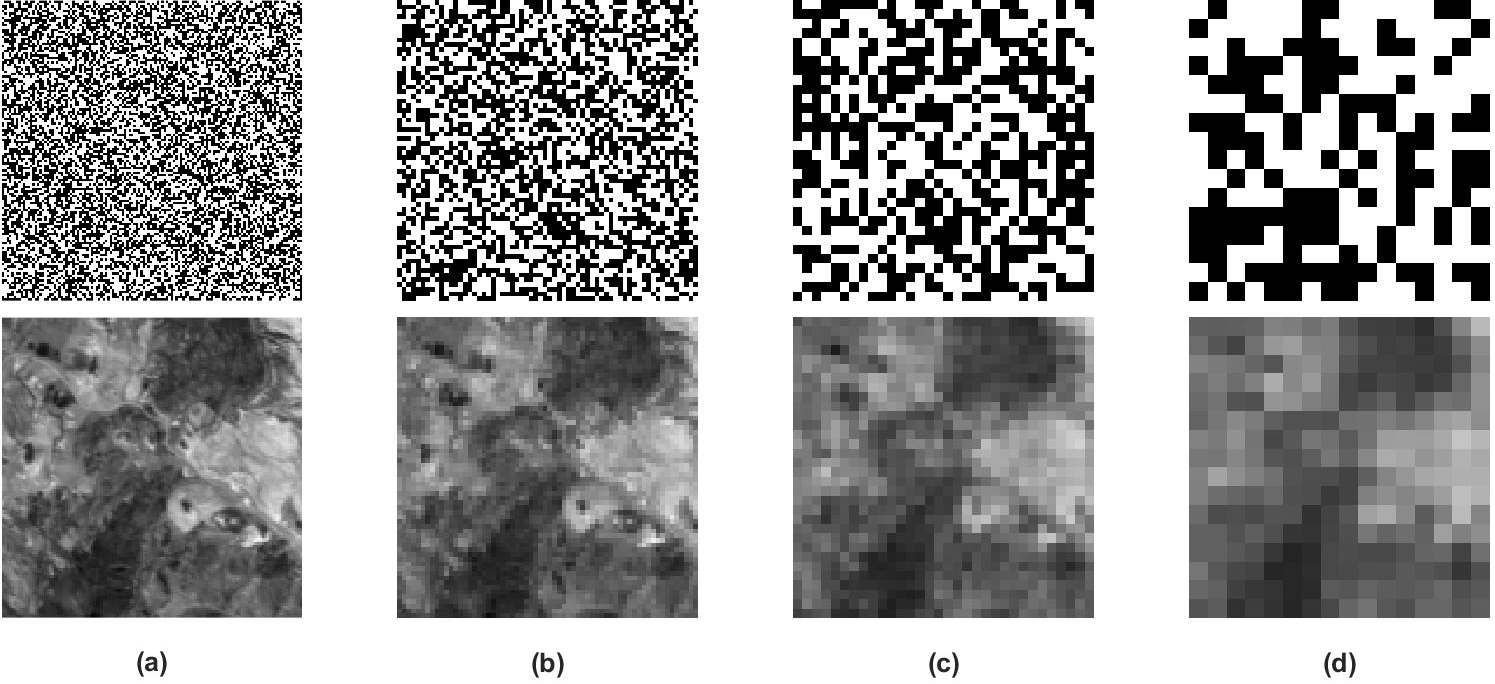}
\caption[Illustration of macro pixels.]{Illustration of macro pixels. \emph{Top panel}: Example of macro pixel measurement masks formed by binning DMD micromirrors. The macro pixels are randomly assigned 0 or 1 values. \emph{Bottom panel}: Macro pixel representation of the Cuprite Vis dataset obtained by replacing each macro pixel block with the mean of the corresponding block from the original image. Macro pixel size (a) $1 \times 1$ (Original), (b) $2 \times 2$, (c) $4 \times 4$, and  (d) $8 \times 8$.}
\label{fig:macropixel}
\end{figure}

The process of binning DMD micromirrors to form macro pixels was discussed in the previous section. Each macro pixel sums the incoming radiation from its field of view and the reciprocal of the size of the macro pixel acts as a cut-off for the spatial frequencies acquired by the measurement masks of that macro pixel size. By reducing the macro pixel size one acquires higher spatial frequencies. The macro pixel size of $1\times1$ defines the native resolution of the DMD. Examples of DMD mirrors binned by different macro pixel size are shown in Fig.~\ref{fig:macropixel}. This method of multi-scale random-macro-pixel sampling will be used to acquire RoIs at different spatial resolutions.

\subsection{Multi-scale Multi-level Sampling with Walsh Transform}

\begin{figure}[!t]
\centering
\includegraphics[width=\columnwidth, height = 5cm, keepaspectratio]{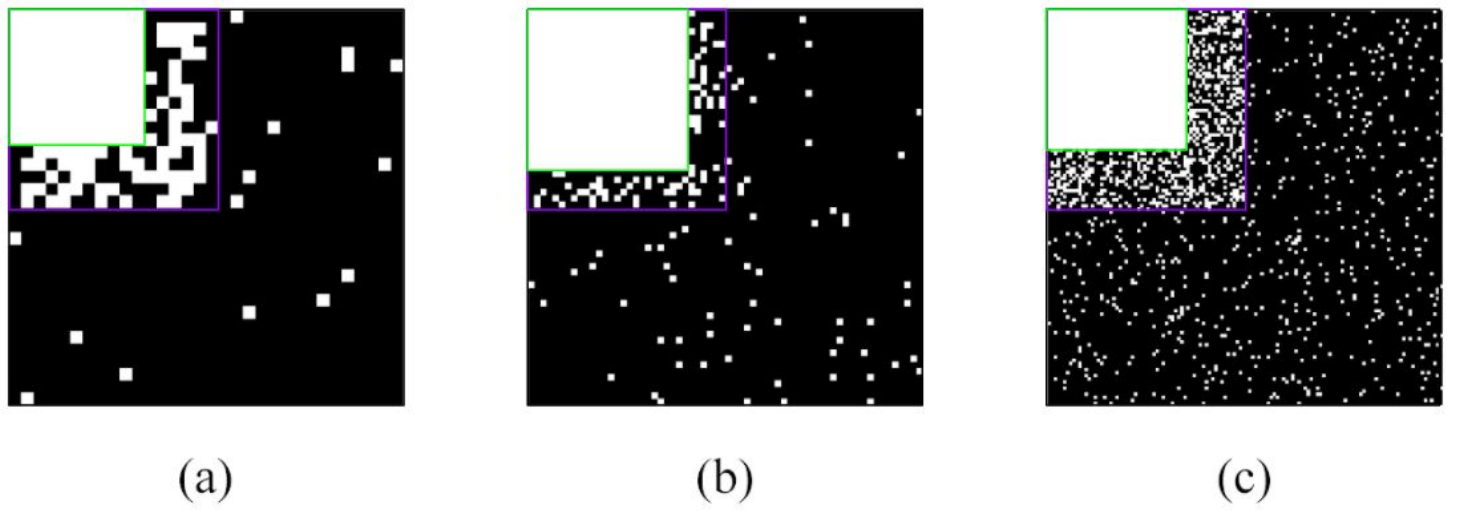} 
\caption{Walsh transform multi-level sampling maps for RoI of size (a) $32 \times 32$, (b) $64 \times 64$, and (c) $128 \times 128$. Top left corner denotes the DC frequency. Number of measurements for each size is 20\% of the total number of pixels in the RoI. The Walsh frequency grids are scaled to the same size for visualisation.}
\label{fig:samplingMaps}
\end{figure}

An alternate way to sample in a multi-scale manner is to use structured matrices like the Walsh transform. Structured matrices have been used in compressed sensing acquisition as they can be implemented on bi-stable DMDs, and have fast forward and inverse transform implementations~\cite{donohoExtension, candesRomberg2}. Another advantage of using Walsh transforms and other structured matrices is that their multi-scale decomposition can be adapted to best leverage the sparsity structure of the signal to be acquired. Multi-level sampling was suggested in~\cite{donohoExtension} and was further developed in~\cite{Lustig_sparseMRI, asymptoticSparsity, roman2014asymptotic}. In multi-level sampling the frequency space of the measurement matrix is divided into a number of levels or regions based on its coherence structure and the number of measurements are assigned to these levels in an asymptotically decreasing manner. The Transform Point Spread Function (TPSF)~\cite{Lustig_sparseMRI} is used as a metric to generate sampling maps for different sized RoIs. In this work, square regions (as shown in Fig.~\ref{fig:samplingMaps}) were chosen instead of the regular circular regions as they allowed easier decomposition of the frequency space into multiple levels. The Walsh frequency grid for each RoI is divided into three levels corresponding to low, mid, and high spatial frequencies. To perform multi-scale sampling we start from the low frequency region followed by level-wise spatial Walsh frequency measurements to achieve a step-by-step increase in resolution. The term multi-scale is a misnomer for multi-level sampling as multi-level sampling itself is defined in a multi-scale manner; however, we use it in the current context to make the stepwise multi-scale acquisition of the scene explicit. Sampling maps for three different size of RoIs are shown in Fig.~\ref{fig:samplingMaps}. The spatial frequency regions may be divided into more than three regions depending upon the application at hand. More information about the number of measurements performed in each spatial frequency region is provided in Sec.~\ref{sec:results}. %\showthe\font

\subsection{Compressed Sensing Algorithms}
\label{sec:csAlgos}

To complete the discussion on acquisition methodology the compressed sensing reconstruction algorithms used are discussed. Two algorithms are used, the Analysis BPDN~\cite{majumdarPriors, prior2} and the Analysis+TV~\cite{majumdarPriors, Lustig_sparseMRI}. The Analysis BPDN can be written as 
\begin{mini}
{x}{\|Wx\|_1 + \gamma \|y - Ax\|_2^2 ,}{\label{reviewEq12}}{ }
\end{mini}
where $A$ is the measurement matrix with normalized columns, $W$ is the sparsity basis, $y$ is the vector of measurements, $x$ is the signal to be recovered, and $\gamma$ is the regularisation parameter. For a signal $ x = \left[ x_1, x_2, \ldots, x_N \right]^T$  of length $N$, the $l_1$ norm is defined as $\|x\|_1 = \sum_{i \in N} |x_i|$ and the $l_2$ norm is defined as $\|x\|_2 = \sqrt{\sum_{i \in N} |x_i|^2}$. Algorithm~(\ref{reviewEq12}) is used for reconstruction of the low resolution image in the first step of the acquisition process with 2D-DCT as the sparsity basis.

The Analysis+TV algorithm
\begin{mini}
{x}{\beta_1 \|Wx\|_1 + \beta_2 \|x\|_{TV}}{\label{reviewEq15}}{ }
\addConstraint{\|y-Ax\|_2^2}{\leq \eta,}
\end{mini} adds a Total Variation (TV) regularisation term to the Analysis BPDN. The TV norm for images is calculated along the horizontal and vertical directions~\cite{tfocs},~$\|x\|_{TV} = \sum_{i,j \in N}\sqrt{|x_{i+1,j} - x_{i,j}|^2 + |x_{i,j+1} - x_{i,j}|^2}$. Such a model is used when the smoothness, promoted by the TV norm, and sparsity constraints, promoted by the $W$ basis, are required simultaneously. The terms $\beta_1$ and $\beta_2$ are used to balance the contributions of the two prior terms, and $\eta$ is used to control the size of the noise.

In this section the camera model and the methods for step-by-step increase in the acquisition of spatial frequencies were discussed. The next section motivates and derives the information change estimator, which is used to develop the proposed RoI prioritisation and acquisition algorithm.

\section{RoI Prioritised Sampling for Constrained Systems}
\label{sec:RI}

\subsection{Refinement Indicator} 
After the ``RoI detection and segmentation" step in the proposed acquisition scheme shown in Fig.~\ref{fig:basicAlgo}, we must distribute the limited measurement budget among the RoIs based on their information content or, in other words, prioritise the RoIs for sequential acquisition. Let $x \in \R^N$ be the RoI to be acquired at the finest resolution. Let us start by acquiring some coarse measurements. The system of equations for the coarse measurements is \begin{align}
y^C = A^C x \ + \ \eta^C,
\label{riEq1} 
\end{align}
where $A^C  = (a_i^C)_{i = 1,...,m} \in \R^N$ is the coarse resolution measurement matrix, the coarse measurements are $y^C$, and $\eta^C$ is an additive noise component. Denoting the solution to the regularised problem as $x^C$, we have
\begin{argmini}
{z}{\frac{1}{2}\|A^Cz - y^C\|_2^2 \ + \ \alpha J(z),}{\label{riEq2}}{x^C = }
\end{argmini}
where $J$ is a regularisation functional like total variation or a wavelet norm and $\alpha \ge 0$ is a regularisation parameter. The optimality condition for~(\ref{riEq2})~\cite{boydSubgradients, Bertsekas} is given by 
\begin{align}
    (A^C)^*(A^Cx^C - y^C) \ + \  \alpha p^C = 0, \ p^C \in \partial J(x^C) ,
    \label{riEq3}
\end{align}where $p^C$ is a subgradient ~\cite{Bertsekas} of $J$ at $x^C$ and $\partial J(x^C)$ denotes the subdifferential~\cite{Bertsekas} of $J$ at $x^C$. The adjoint of $\Ac$ is denoted by $\Acs$. Let us ignore the noise term for the moment so that $y^C = A^Cx$. To improve the resolution we acquire extra measurements with a refinement matrix $B$. This matrix is composed of sampling patterns that acquire finer spatial frequencies than $A^C$. For example, if measurements in $A^C$ have a macro pixel size of $8 \times 8$ then $B$ would contain random measurements with macro pixel size of $4 \times 4$. The measurement matrix with the additional refined measurements can be written as
\begin{align}
A^R =  \left(\begin{array}{c} A^C \\ B \end{array}\right).
\label{riEq4}
\end{align}The refined measurements are given by $y^R = A^Rx$. Again the refined solution $\xr$ satisfies the optimality conditions
\begin{align}
    (A^R)^*(A^Rx^R - y^R) \ + \ \alpha p^R \ =\  0, \ p^R  \in  \partial J(x^R), 
    \label{riEq5}
\end{align} where $p^R$ is a subgradient of $J$ at $x^R$ and $\partial J(x^R)$ is the subdifferential of $J$ at $x^R$. To be explicit the refined solution $x^R$ contains spatial frequencies unique to the acquisition through $B$ along with those acquired previously through $A^C$. To estimate the change in information of the RoI when acquired at different resolutions we would like to derive an {\em a-posteriori} error estimate between the coarse solution, $x^C$, and the refined solution, $x^R$, without the evaluation of the latter. This would quantify the information change across scales for the RoI. We follow the procedure for calculating error estimates in~\cite{burger2016, burger2018}. From~(\ref{riEq4}) we get,  

\begin{align}
\Ars\Ar = \Acs \Ac + B^*B.
\label{riEq7}
\end{align} Subtracting~(\ref{riEq5}) from~(\ref{riEq3}) we get, 
\begin{equation}
\begin{aligned}
\Acs \Ac\xc \ - \  \Acs \yc \  + \  \alpha(\pc - \pr)  \ \\
- \ \Ars \Ar \xr \ + \  \Ars\yr = 0.
\label{riEq8}
\end{aligned}
\end{equation} Using $\yc = \Ac x$, $\yr = \Ar x$, and the value of $\Acs \Ac$ from~(\ref{riEq7}) we get,
\begin{align}
\Ars\Ar(\xc - \xr) \ + \ \alpha(\pc - \pr) = \Bs B(\xc - x).
\label{riEq9}
\end{align}Taking a scalar product of the above with $(\xc - \xr)$ yields

\begin{multline}
\norm{\Ar(\xc - \xr)}  \ +  \ \alpha D_J(\xc, \xr) \\ 
 = \langle B(\xc - x), B(\xc - \xr)\rangle,
\label{riEq10}
\end{multline}where $D_J = \langle \pc - \pr, \xc - \xr \rangle$ is the symmetric Bregman distance~\cite{burger2016}. Applying Young's Inequality, $\langle u, v\rangle \leq  \frac{1}{2}\norm{u} + \frac{1}{2}\norm{v}$, to the right hand side of~(\ref{riEq10}) yields

%\begin{equation}
%\begin{aligned}
\begin{multline}
\norm{\Ar(\xc - \xr)} \ + \ \alpha D_J(\xc,  \xr) \\ \leq \frac{1}{2} \norm{B(\xc - x)} \ + \ \frac{1}{2}\norm{B(\xc - \xr)}. 
\label{riEq11}
\end{multline}Using the fact that, $\norm{Bv} \leq \norm{Bv} + \norm{\Ac v}
 = \norm{\Ar v}$ for any $v$,~(\ref{riEq11}) becomes
%\begin{equation}
%\begin{aligned}
\begin{multline}
\frac{1}{2}\norm{\Ar(\xc - \xr)} \ + \ \alpha D_J(\xc, \xr) \\ \leq \frac{1}{2} \norm{y^B - B\xc},
\label{riEq12}
\end{multline}
%\end{aligned}
%\end{equation}
where $y^B = Bx$ are the novel measurements. The second term on the left-hand side, $D_J(\xc, \xr)$, is the error between the two solutions at coarse and fine resolutions with respect to the functional $J$.  For $J = \|.\|_1$, the symmetric Bregman distance, $D_J(\xc, \xr)  = 2 \Sigma_{sign(\xc_i) \neq sign(\xr_i) } \vert \xr_i - \xc_i \vert$, measures the deviation between entries of the two solutions that differ in their signs~\cite{burger2018, burger2016}. For a wavelet norm, % as in~(\ref{reviewEq12})
the symmetric Bregman distance will measure the deviation in the sparsity pattern of the wavelet representations of the two solutions~\cite{burger2007}. If the entries of the wavelet representations have the same sign,~$D_J(\xc,\xr) = 0$ and it is non-negative if the entries differ in their signs. Further, the first term on the left-hand side of~(\ref{riEq12}) measures how well we estimate the magnitudes of the entries of each of the two solutions, the residual between the two solutions. The behaviour of the Bregman distance with respect to the parameter~$\alpha$ and its asymptotic are well understood and not further discussed here, see~\cite{burger2016}. The right-hand side of~(\ref{riEq12}) is an a-posteriori estimator for the change between~$\xc$ and~$\xr$ given only the coarse solution~$\xc$. It is an a-posteriori error estimator in the \emph{computational} sense, it can be computed without solving a fine (the refined) scale problem. It is not an a-posteriori in the \emph{measurement} sense, since we need to collect at least some fine scale measurements (refined measurements). Taking into consideration that $B$ is random, we can take expectation on both sides of~(\ref{riEq12}) 

\begin{equation}
\begin{aligned}
\frac{1}{2}\mathbb{E}[\, \|A^R(\xc - \xr)\|_2^2 ]\, \ + \ \alpha  \mathbb{E}[\,D_{J}(\xc,\xr)]\, \\ \leq  \frac{1}{2} \mathbb{E}[\,\|\yb - B\xc \|_2^2]\,.
    \label{riEq13}
\end{aligned} 
\end{equation}
%\begin{equation}
%\begin{aligned}
% \mathbb{E}[\,D_{J}(\xc,\xr)]\,  \leq  \frac{1}{2 \alpha} \mathbb{E}[\,\|\yb - B\xc \|_2^2]\,.
%    \label{riEq13}
%\end{aligned} 
%\end{equation}
This shows that in expectation the maximum error between $\xc$ and $\xr$ occurs when the expected deviation between $\yb$ and $B\xc$ is maximal. Thus, the Refinement Indicator (RI)  can be defined as $\|\yb - B\xc\|_2^2$. The RI can be calculated for each RoI in the scene and the RoI with the largest value of RI can be refined. Explicitly, to refine an RoI is to calculate the higher spatial resolution solution from the measurements $y^R$ obtained through the measurement matrix $A^R$. Intuitively, for a flat region, $\xc$ will be a good approximation of the underlying region, therefore, the novel measurements, $\yb$, and the \textit{simulated} measurements obtained through $B$ by assuming $\xc$ to be the groundtruth; i.e. $B\xc$, will be similar and the value of RI will be small. On the other hand, if the underlying region is textured, $\xc$ will be a poor approximation to the region and the novel measurements and \textit{simulated} measurements will differ leading to a larger value of RI. Thus, the RI can be used to prioritise RoIs for acquisition. 

\subsection{Adaptive Refinement of Measurements} 

\begin{figure}[!t]
\centering
\includegraphics[width=\columnwidth, keepaspectratio]{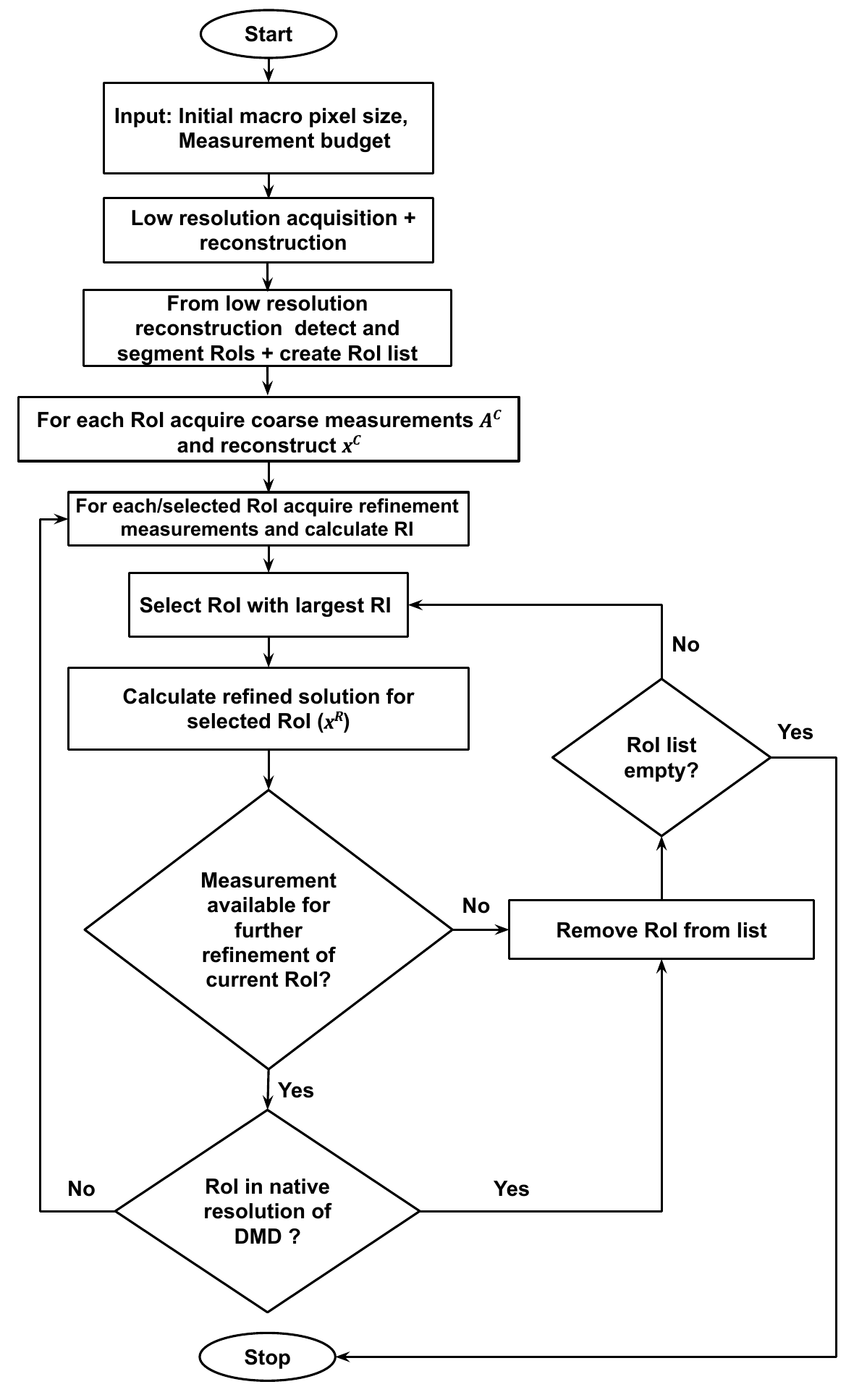}
\caption{RPS Algorithm for RoI prioritisation using RI in measurement-constrained autonomous exploration systems.}
\label{algoRI}
\end{figure}

The RI developed in the previous section is used to develop an acquisition algorithm for measurement-constrained systems. The three step sampling procedure discussed in~\cite{MCS} and further developed in Sec.~\ref{sec:cameraModel} is used as the basic framework for acquisition. The inputs to the algorithm are the initial macro pixel size of the low resolution acquisition in the first step and a fixed number of measurements, called the measurement budget. After the ``Low resolution acquisition and reconstruction" and ``RoI detection and segmentation" steps we obtain the acquisition masks for each RoI. Coarse measurements for each RoI are performed and the coarse solution $\xc$ is calculated. These coarse measurements differ from the ones acquired in the ``Low resolution acquisition and reconstruction" step because the low resolution measurements are multiplexed measurements from the entire scene and recovering individual RoIs from multiplexed measurements is not possible. Additional random measurements $\yb$ refining the previous coarse measurements are acquired for each RoI. The RI is calculated for all the RoIs, the RoI with the largest RI is chosen and the refined solution, $\xr$, is calculated for that RoI and it is set as the new coarse level for the chosen RoI. New refinement measurements are acquired at a higher spatial resolution for the current refined RoI and the new RI is calculated from the new refined measurements. RI values for the other RoIs are brought forward and again the RoI with the largest value of RI is chosen for refinement. This process continues until all the RoIs are in the native resolution of the DMD or until the measurement budget is exhausted. This results in prioritisation of  RoIs in terms of the information content change across spatial resolutions. The limited measurements are spent on RoIs with greater change in information content. If measurements are not available for refining a particular RoI then it is removed from the prioritisation list and the RIs of the remaining RoIs are compared for refinement. The complete algorithm is shown in Fig.~\ref{algoRI} and is named RoI~Prioritised~Sampling~(RPS).

\subsection{Discussion}

One can consider a na\"ive approach in which one acquires RoIs at the coarse resolution, performs coarse reconstructions and acquires further refined measurements for all the RoIs. One can reconstruct all the RoIs at this refined resolution and then select the RoI that produced maximum change in information content between the coarse and the refined image of the RoI for further refinement, expecting the trend of increase in information content over scales to continue. However, the decision on which RoI should be refined further can only be made \textit{after the reconstruction of} all RoIs at the refined resolution. The extra reconstructions may be expensive for a constrained system. This problem is alleviated with the RPS, as RI allows us to make the same decision \textit{without} calculation of the refined solution for \textit{all} RoIs. In limited measurement settings, like remote exploration, acquired refined measurements can be sent to a ground-station where resources would be available for reconstructions. Thus, we focus specifically on a measurement-constrained setting. Also, the RPS is concerned with utilising the limited measurement budget optimally and not specifically with reducing the number of measurements required to acquire the scene/RoI.

\section{Results}
\label{sec:results}

\begin{figure}[!t]
\centering
\includegraphics[width=\columnwidth, keepaspectratio]{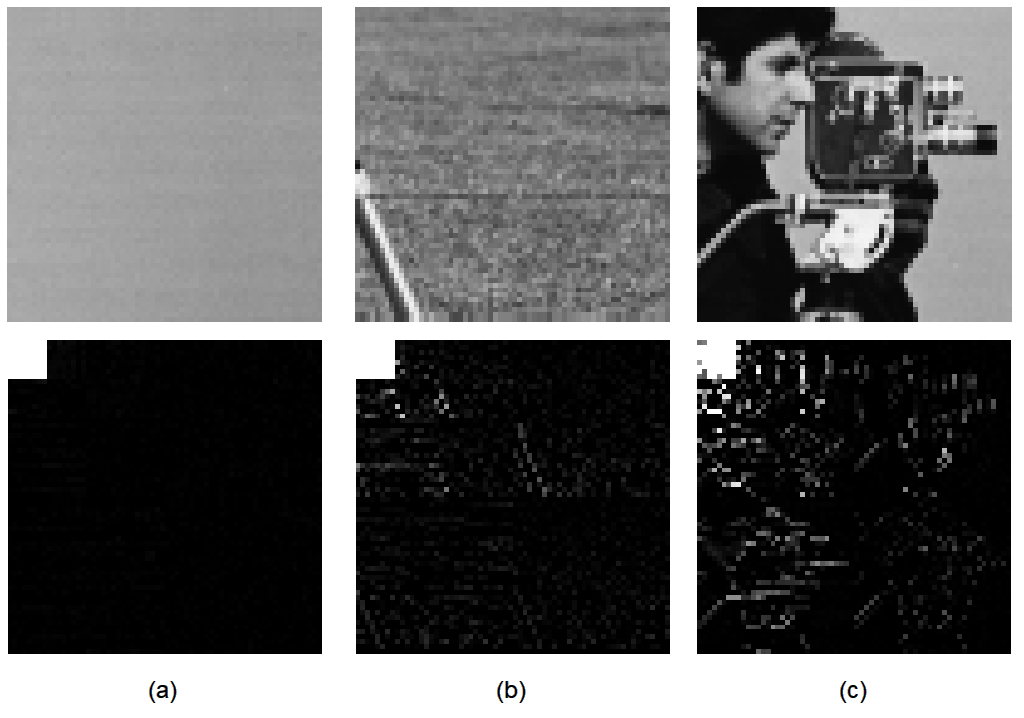}
\caption[Second example of flat and textured regions in an image.]{\emph{Top Panel}: Three different RoIs from the cameraman image with different amount of textures. \emph{Bottom Panel}: The three-level Haar wavelet decomposition of the corresponding RoIs. (a) RoI with minimal textures (flat region), (b) RoI with some textures, and (c) RoI with rich textures.}
\label{figCameramanRoI}
\end{figure}

To demonstrate that the RI does in fact quantify information content change at different spatial scales, three different regions of the cameraman image were selected with varying amount of textures. The RoIs and their three-level Haar wavelet decompositions are shown in Fig.~\ref{figCameramanRoI}. Random-macro-pixel measurements with $0/1$ entries are performed on each RoI starting from a macro pixel size of $8 \times 8$ and the $8 \times 8$ reconstruction is calculated. Thereafter, a refinement matrix $B$ is generated by using random-macro-pixel sampling with a macro pixel size of $4 \times 4$ and novel measurements are generated for each RoI. \textit{Simulated} measurements are also generated using the reconstruction of the $8 \times 8$ macro pixel acquisition and the refinement matrix $B$. The two measurements are subtracted and the $l_2$ norm of the error vector is calculated to form the RI. The process is repeated for macro pixel size of $2 \times 2$.  The RI values for the RoIs are recorded in Table~\ref{tableRIcameraman}. The RI for spatial resolution $8 \times 8$ denotes the information change in a RoI when measurements are refined from a resolution of $8 \times 8$ to $4 \times 4$. The last row in the table defines the RI value of refining a RoI from $2 \times 2$ to $1 \times 1$, the native resolution of the DMD, and there is no RI for spatial resolution of $1 \times 1$ macro pixel. The largest change in the value of the RI is consistently obtained for the RoI that has the largest number of high frequency components, i.e., for  RoI~(c) in Fig.~\ref{figCameramanRoI}. The smallest change in RI value is observed for RoI~(a) that has the smallest number of high frequency components. Thus, the RI estimates the change in information content across spatial scales and is a relevant metric that can be used to prioritise RoIs for acquisition in measurement-constrained systems.
 
Further, the RPS algorithm developed in the previous section was tested on images derived from different sensors. The experiments are designed to reflect remote exploration scenarios from Earth observation applications and extra-terrestrial exploration.  After a short overview of the datasets the results of applying RPS on the datasets will be discussed and the efficacy of the proposed method in prioritising RoIs in measurement-constrained systems will be verified. The algorithm was implemented in MATLAB 2018a and simulations were performed on a Windows system with 16 GB RAM and an Intel(R) i7-6700 CPU @ 3.4GHz. Code for reproducing the results can be found at  https://github.com/protim1191/RoI-Prioritised-Sampling.git.

\begin{table}
\caption{RI values for RoIs from Fig.~\ref{figCameramanRoI} calculated at spatial resolutions of $8 \times 8$, $4 \times 4$, and $2 \times 2$ macro pixels. Random-macro-pixel measurements with $0/1$ entries were used for each spatial resolution. The number of refined measurements at each macro pixel resolution for each RoI is~409.}
\begin{center}
\begin{tabularx}{\columnwidth}{|>{\centering}X|>{\centering} X|>{\centering} X|>{\centering\arraybackslash} X| }
\hline
	Spatial Resolution &  RI for RoI Fig.~\ref{figCameramanRoI}(a) &  RI for RoI  Fig.~\ref{figCameramanRoI}(b) & RI for RoI  Fig.~\ref{figCameramanRoI}(c)\\	
	\hline
	$8 \times 8$ &  9.74 &  97.91& 218.11 \\
	\hline
	$4 \times 4$ & 5.79 & 61.58 & 110.37\\
	\hline
	$2 \times 2$ & 5.10 & 39.34 & 46.99\\
	\hline
\end{tabularx}
\label{tableRIcameraman}
\end{center}
\end{table}

\subsection{Datasets}

\subsubsection{Cuprite}
The Cuprite geological dataset is a snapshot of the cuprite ores in the state of Nevada, USA. The hyperspectral datacube was acquired by the AVIRIS~\cite{avirisWebsite} sensor that collects data in the wavelength range of 400 to 2500 nm with a nominal channel bandwidth of 10 nm. The ground sampling area is 20~m$^2$ and the radiometric resolution is 16 bits. The datacube is divided into two, the visible and the infrared. The visible section consists of wavelengths 400 to 800 nm and is called the ``Cuprite Vis" dataset; the infrared dataset consists of wavelengths from 900 to 2500 nm and is called the ``Cuprite~IR" dataset. 

\subsubsection{Gulf of Mexico}

The Gulf of Mexico dataset is a part of baseline datasets provided by SpecTIR~\cite{specTIR}. The sensor acquires data in the wavelength range of 395 to 2450 nm with a nominal channel bandwidth of $5$ nm. The ground sampling distance is $2$ m. As before the datacube is divided into visible and infrared regions. The visible region consists of wavelengths from 400 to 800 nm and is called the ``Gulf Vis" dataset. The infrared region consists of wavelengths from 900 to 2400~nm and is called ``Gulf IR" dataset. Only the Gulf Vis dataset is used for experiments as the spatial variations in the two datasets are similar.

\subsubsection{University of Pavia}
This scene was acquired by the ROSIS sensor~\cite{rosisPaper} during a flight campaign over Pavia in northern Italy. The detector is sensitive to wavelengths ranging from 430 to 860 nm with a nominal bandwidth of 5 nm. Certain pixels did not contain any information and were  discarded. The entire dataset is considered to be a part of the visible wavelengths and is called ``Pavia" in this work.

\subsubsection{MSL ChemCam Images}
\begin{figure}[!t]
\centering
\includegraphics[width=\columnwidth,  keepaspectratio]{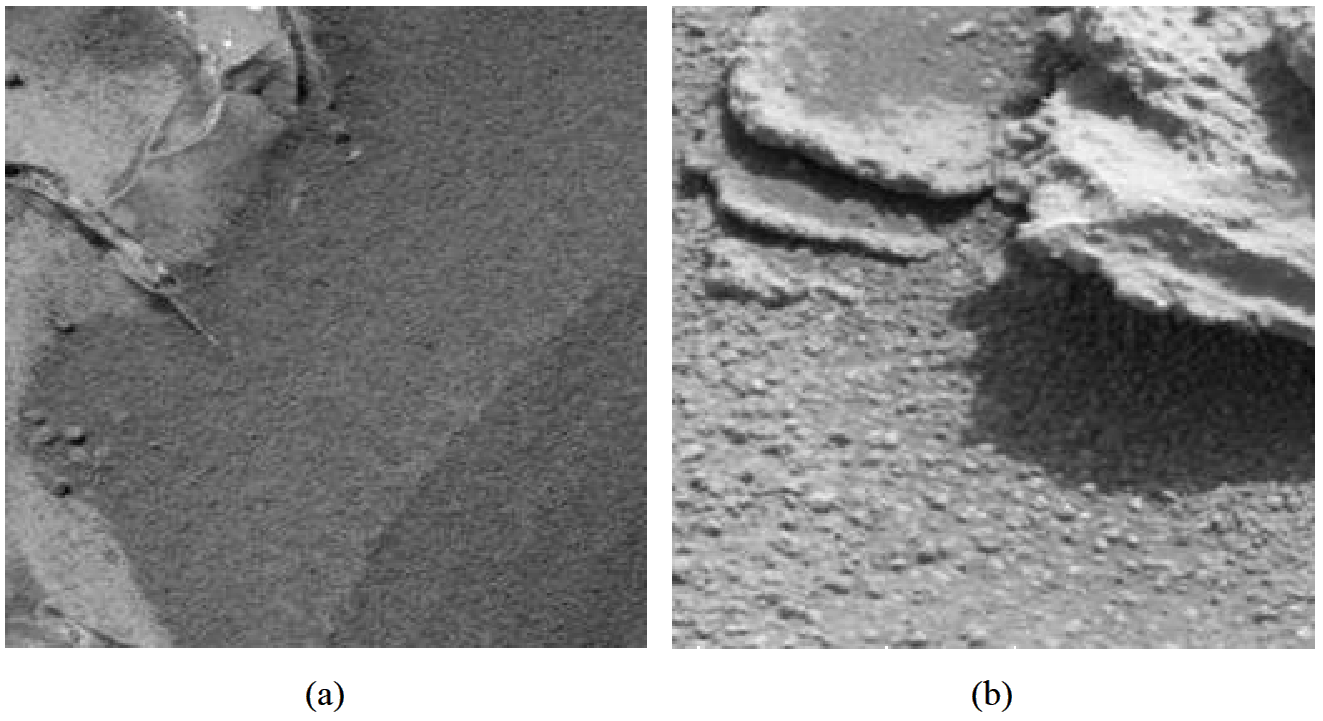}
\caption{Datasets from the Mars Science Laboratory on the Curiosity rover used for evaluation of RPS. (a) Mars1 and  (b) Mars2.}
\label{fig:datasetsMars}
\end{figure}

Two images from the Mars Science Laboratory (MSL)~\cite{curiosityMSLImages} on the Curiosity rover~$\textit{0133\_crc\_ccam01133\_rowatt.png}$ and $\textit{0121\_crc\_ccam01121\_stanbridge.png}$ have been used to test the RPS algorithm. The test images are called ``Mars1" and ``Mars2" respectively. These are shown in Fig.~\ref{fig:datasetsMars}.

\begin{figure}[!t]
\centering
\includegraphics[width=\columnwidth, height = 10cm, keepaspectratio]{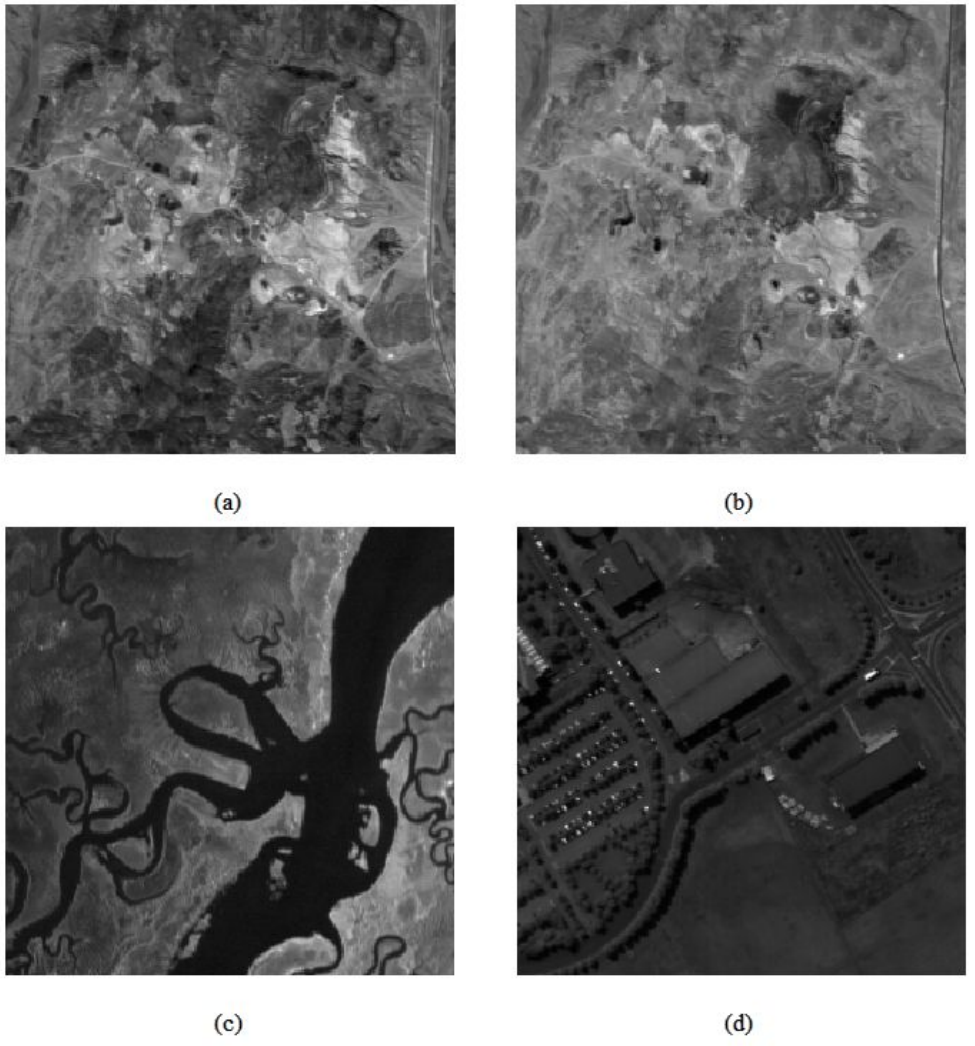}
\caption{Remote sensing datasets used for evaluation of RPS. (a) Cuprite Vis dataset, (b) Cuprite IR dataset, (c) Gulf Vis dataset, and (d) Pavia dataset.}
\label{fig:datasets}
\end{figure}

\subsection{Preprocessing of Datasets and Algorithm Parameters}
For all the datasets a $256 \times 256$ section was used. After removing corrupted bands, pixel values in each spectral band were converted to be between 0 and 1 by using min-max normalisation. A single channel image was formed from the normalised spectral image by averaging across the spectral dimension of each pixel. The SPC architecture described in Sec.~\ref{sec:spc} was used to simulate the acquisition process. The measurement budget was different for each dataset and is specified along with their respective results.  The number of low resolution measurements in the first step was fixed at 1000 and the macro pixel size used was $8 \times 8$. The number of regions selected during the ``RoI detection and segmentation" procedure was fixed at 10. Regions detected within 1 pixel radius of each other were merged. Thus, the final number of RoIs available for refinement is different for each dataset. The acquisition and recovery of each RoI is independent of the other RoIs, therefore, the overlap regions common to two RoIs is sampled twice at the same resolution. However, due to budget considerations if only one of the RoIs can be acquired, then the overlap region will be at higher resolution for the unacquired RoI as well. As the algorithm keeps track of the resolution of each RoI, further visualization or analysis of the overlap region can be restricted to the higher resolution RoI.  The number of measurements used to calculate the RI was fixed at 10\% of the total number of pixels in the RoI for random-macro-pixel sampling and at 20\% for multi-scale multi-level sampling.  The RoI with the largest value of RI is chosen for refinement at each step. 
\begin{figure}[!t]
\centering
\includegraphics[width=\columnwidth, keepaspectratio]{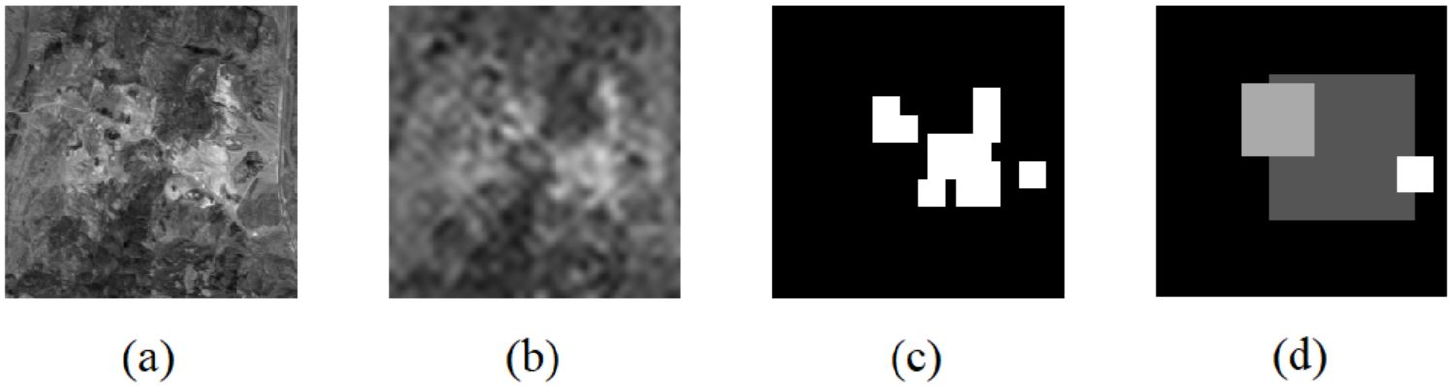}
\caption[First two steps of MCS.]{Low resolution reconstruction and RoI detection and segmentation process for Cuprite Vis dataset. (a) Original, (b) low resolution reconstruction from 1000 measurements, (c)~detected and segmented RoIs, and (d) merged RoIs. The three gray-levels represent three different RoIs for which the RI will be calculated.}
\label{figFirstSteps}
\end{figure}

\begin{figure}[!t]
\centering
\includegraphics[width=\columnwidth, keepaspectratio]{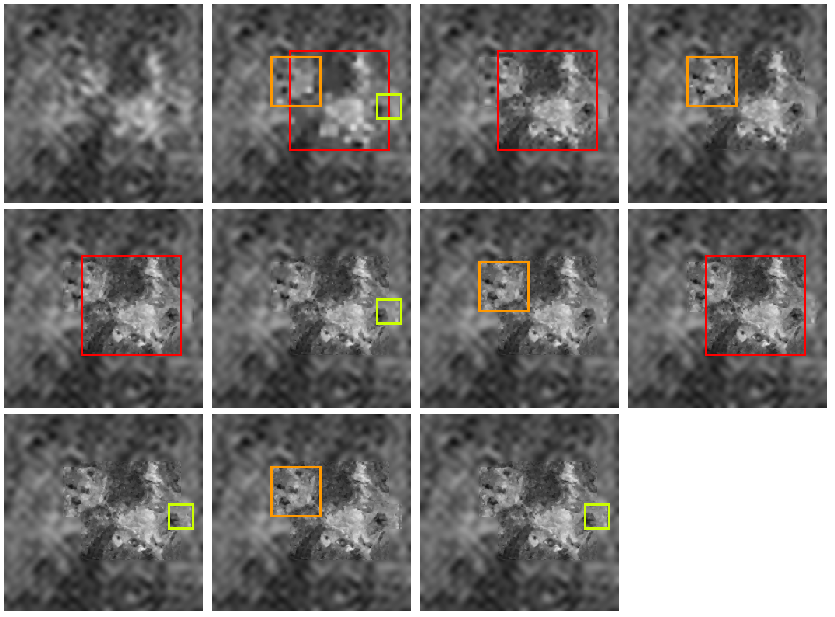}
\caption[Prioritisation of RoIs with RI for Cuprite Vis dataset.]{Prioritisation of RoIs with RI for Cuprite Vis dataset. The total number of measurements is 9600 (14.6\% of the total number of pixels).}
\label{figCupVisRI}
\end{figure} Upon experimentation algorithm~(\ref{reviewEq12}) was found to be better at the initial low resolution reconstruction of the entire scene with 2D-DCT as the sparsity basis. For subsequent reconstruction of individual RoIs using RPS, algorithm~(\ref{reviewEq15}) was used with Daubechies-8 wavelets~\cite{daubechiesTenLectures} as the sparsity basis. The algorithms were implemented in the TFOCS~\cite{tfocs} toolbox. The maximum number of iterations was fixed at 20000, the algorithm used was `N07', continuation was set to `True' and the number of continuation iterations were set to~3, the regularisation parameters $\beta_1$ and $\beta_2$ from~(\ref{reviewEq15}) were set to~1 and~0.4 respectively for random-macro-pixel sampling and $\beta_1 = 1$ and $\beta_2 = 0.6$ for Walsh measurements. The measurement and sparsity operators were implemented in the SPOT toolbox~\cite{spot}. The results of ``Low resolution acquisition and reconstruction" and ``RoI detection and segmentation" process for Cuprite Vis dataset is shown in Fig.~\ref{figFirstSteps}. Results after segmentation are discussed next.

\begin{table}
\caption[Evolution of RI for Cuprite Vis dataset.]{Evolution of RI for Cuprite Vis dataset. The highlighted RoI indicates the one selected at each iteration. The refined macro pixel resolution of the selected RoI is tabulated. The total number of measurements is 9600.}
\begin{center}
\begin{tabularx}{\columnwidth}{|>{\centering}X|>{\centering}X|>{\centering}X|>{\centering}X|>{\centering}X|>{\centering\arraybackslash}X|}
\hline
%	& \multicolumn{3}{c|}{Refinement Indicator}  & & \\	
& \multicolumn{3}{c|}{RI}  & & \\	
	%\hline
	\cline{2-4}
%	\center{Iteration No.}& RoI 1 (Red) ($128 \times 128$) & RoI 2 (Orange) ($64 \times 64$) & RoI 3 (Yellow) ($32 \times 32$) & Refined Macro Pixel Resolution & Available Measurements \\
Iteration No.&  \makecell{ RoI 1\\ \hspace{-1mm}(Red)\\ \hspace{-1.5mm}($128\! \times\! 128$)} &  \makecell{RoI 2\\ (Orange)\\ ($64 \times 64$)} &  \makecell{RoI 3 \\(Yellow)\\($32 \times 32$) }&  Refined Macro Pixel Resolution &  Available Measurements \\
%	\center{Iteration No.}&RoI 1\newline (Red)\newline($128 \times 128$) & \thead{RoI 2\\ (Orange)\\ ($64 \times 64$)} & \thead{RoI 3\\ (Yellow)\\ ($32 \times 32$)} & Refined Macro Pixel Resolution & Available Measurements \\

	\hline
%1 & \cellcolor{green} $3.7 \times 10^5$ & $2 \times 10^4$& 973.6& $4 \times 4$&4302  \\
1 & \cellcolor{green} 349.30 &134.64 &14.82& $4 \times 4$&4302  \\
	\hline
%2 & \cellcolor{green} $6 \times 10^4$ & $2 \times 10^4$& 973.6& $2 \times 2$& 2664 \\
2 & 56.98 &\cellcolor{green}134.64 & 14.82& $4 \times 4$& 2664 \\
\hline
%3 & $1.4 \times 10^4$ & \cellcolor{green}$2 \times 10^4$& 973.6&$4 \times 4$&1026 \\
3 & \cellcolor{green} 56.98&14.64& 14.82 &$2 \times 2$&2255 \\
\hline
%4 & \cellcolor{green} $1.4 \times 10^4$ & $4.1 \times 10^3$& 973.6 & $1 \times 1$ &617\\
4 & 11.73 & 14.64&\cellcolor{green} 14.82 & $4 \times 4 $ &617\\
\hline
%5 & -  & \cellcolor{green}$ 4.1\times 10^3$& 973.6& $2 \times 2$ &617 \\
5&11.73 & \cellcolor{green}14.64  & 6.57&  $2 \times 2$ &515 \\
\hline
%6 & - & \cellcolor{green}995.6& 973.6&$1 \times 1$&208 \\
6 & \cellcolor{green}11.73&3.36&6.57&$1 \times 1$&106\\
\hline
%7 & - & - & \cellcolor{green}973.6& $4 \times 4$&208 \\
7 & - & 3.36 & \cellcolor{green}6.57& $2 \times 2$&106 \\
\hline
%8 & -&-& \cellcolor{green}284.4& $2 \times 2$&106\\
8 & -&\cellcolor{green}3.36& 0.80& $1 \times 1$&4\\
\hline
%9 & - & - & \cellcolor{green}76.62 &$1 \times 1$&4 \\
9 & - & - & \cellcolor{green}0.8 &$1 \times 1$&4 \\
\hline
10 & -  & - &- &-&4\\
\hline
\end{tabularx}
\label{tableCupVisRI}
\end{center}
\end{table}

\subsection{Results on Remote Sensing Datasets}

This section discusses the results for the remote sensing datasets, i.e., Cuprtie Vis, Cuprite IR, Gulf Vis, and Pavia, shown in Fig.~\ref{fig:datasets}. For the following results multi-scale random-macro-pixel sampling with $0/1$ entries was used. The results of applying the RPS algorithm on the Cuprite Vis dataset is shown in Fig.~\ref{figCupVisRI}. The first image from the left in the top row is the low resolution reconstruction. The second image from the left in the top row shows the coarse reconstruction of all the detected RoIs. Continuing from left to right and top to bottom, the prioritisation of the RoIs and their refined reconstructions with the progress of the RPS is shown. Table~\ref{tableCupVisRI} records the change in the RI values and the choice of RoI made at each iteration of the algorithm along with the number of available measurements at each iteration that decides whether further refinement is possible for a RoI. For the Cuprite Vis dataset a total of 9600 measurements was used for acquisition, which is 14.6\% of the total number of pixels in the scene. The number of low resolution measurements is 1000 and these are used to acquire the entire scene at $8 \times 8$ macro pixel resolution. Three RoIs of size $128 \times 128$, $64 \times 64$, and $32 \times 32$ are selected after the ``RoI detection and segmentation" process (Fig.~\ref{figFirstSteps}). A total of  2149 (1638 + 409 + 102) coarse measurements were used.

Another 2149 measurements were used for  the initial RI values for the three RoIs. These values indicate the change in information content for each RoI when refined to a macro pixel resolution of~$4 \times 4$ from their current~$8 \times 8$ macro pixel resolution. The~$128 \times 128$ sized RoI, marked in red in Fig.~\ref{figCupVisRI}, is selected for refinement as it has the largest RI value. Upon calculation of the refined solution of the selected RoI the number of available measurements is checked to decide whether enough measurements is available for calculation of the new RI for the selected RoI, which would estimate the change in information one could expect when the RoI is refined to a macro pixel resolution of~$2 \times 2$ from the current~$4 \times 4$. As there are~4302 available measurements at iteration~1 and 1638 measurements are required for the selected RoI, refined measurements are performed on the selected RoI and the new RI is calculated. At the second iteration the same sequence of steps is followed, the RoI with the largest value of RI is selected for refinement and the new RI is calculated subject to the availability of measurements. The last RI calculation is done for estimating the change in information when refining from $2 \times 2$ to $1 \times 1$ macro pixel resolution. If the number of available measurements allowed this acquisition then the $1 \times 1$ solution is always calculated, which may also be performed at the ground-station as the measurements for this reconstruction step have already been acquired for the calculation of the RI. Once a RoI is resolved to $1 \times 1$ macro pixel resolution it is replaced with a `-' in the table. This can be seen in iteration~6 and~7 in Table~\ref{tableCupVisRI} where at iteration~6 the red RoI is selected for refinement to a macro pixel resolution of~$1 \times 1$ and replaced with a `-' in iteration~7 but no measurements are utilised as no further refinement is possible for this RoI. As a result, the same number of measurements is available at iteration~7 for RI calculations as at iteration~6.
 
The refinement measurements used to calculate the RI are also used for calculating the refined solution. Table~\ref{tableCupVisRI_roiMetric} records the quality of reconstruction of the selected RoIs at different macro pixel resolutions. The quality of recovery for each RoI improves with acquisition of higher spatial frequency components. Two error metrics are used, namely the Normalised Mean Squared Error (NMSE) and the Structural SIMilarity Index (SSIM)~\cite{ssimPaper}. NMSE between a vector $x$ and its estimate $\hat{x}$ is defined as $\|\hat{x} - x\|_2^2/\|x\|_2^2$. SSIM is a full referential perceptual visual quality metric that considers local luminance, contrast, and structural variance to calculate similarity between the estimate and the reference image. The visual improvements in the reconstructions can be seen in Fig.~\ref{figCupVisRoIs} in the supplementary material.

\begin{table}
\caption{Error metrics for the RoIs for Cuprite Vis dataset at various macro pixel resolution.}
\begin{center}
\newcolumntype{P}[1]{>{\centering\arraybackslash}p{#1}}
\begin{tabularx}{\columnwidth}{|P{1.1cm}|>{\centering}X|>{\centering}X|>{\centering}X|>{\centering}X|>{\centering}X|>{\centering\arraybackslash}X|}
\hline
Resolution of RoI	& \multicolumn{2}{>{\centering\setlength\hsize{2\hsize} }X|}{\thead{RoI 1 \\ (Red) \\ ($128 \times 128$)}}& \multicolumn{2}{>{\centering\setlength\hsize{2\hsize} }X|}{\thead{RoI 2 \\ (Orange) \\($64 \times 64$)}}& \multicolumn{2}{>{\centering\setlength\hsize{2\hsize} }X|}{\thead{ RoI 3 \\ (Yellow)\\ ($32 \times 32$)}}\\	
%	\hline
\cline{2-7}
 & NMSE & SSIM & NMSE & SSIM & NMSE & SSIM  \\
	\hline
$8 \times 8$& 0.032 &0.40 &0.041&0.36 & 0.024& 0.38 \\
\hline
$4 \times 4$&0.015&0.64 &0.016&0.64 &0.015 &0.56 \\
\hline
$2 \times 2$&0.008 & 0.79& 0.009&0.79 & 0.006& 0.78\\
\hline
$1 \times 1$& 0.007& 0.83 & 0.007& 0.83 &0.005 &0.82 \\
\hline
\end{tabularx}
\label{tableCupVisRI_roiMetric}
\end{center}
\end{table}

\begin{figure}[!t]
\centering
\includegraphics[width=\columnwidth, keepaspectratio]{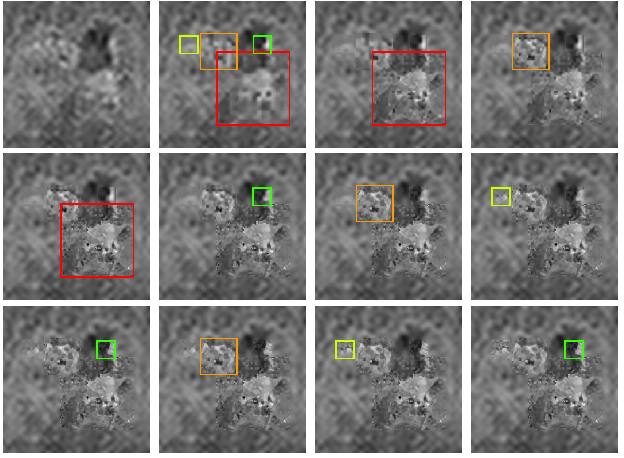}
\caption[Prioritisation of RoIs with RI for Cuprite IR dataset.]{Prioritisation of RoIs with RI for Cuprite IR dataset. The total number of measurements is 8060 (12.3\% of the total number of pixels). }
\label{figCupIRRI_limitedBudget}
\end{figure}

Results of applying RPS on the Cuprite IR dataset are shown in Fig.~\ref{figCupIRRI_limitedBudget}. As shown in Table~\ref{tableCupIRRI_limitedMeasurement}, the online distribution of measurements is more evident here. After coarse reconstruction the RI values for each RoI is calculated. The RoI with the largest RI value, marked in red, is selected for refinement. The available number of measurements allows to calculate the RI for a macro pixel resolution of~$2 \times 2$. In the second iteration, the RoI of size~$64 \times 64$, marked in orange in Fig.~\ref{figCupIRRI_limitedBudget}, has the largest value of RI and is refined to a macro pixel resolution of~$4 \times 4$. In the third iteration, the red RoI has the largest RI value and is refined to a macro pixel resolution of~$2 \times 2$. However, there is not enough measurements to calculate the RI for a refinement to a macro pixel resolution of~$1 \times 1$. Therefore, the red RoI is removed from the list of RoIs due to unavailability of measurements and replaced with a~`-'. The same number of measurements is available at iteration~4 as at iteration~3 because no refinement measurements were used in iteration~3. Thereafter, the process of selecting the RoI with the largest value of RI, calculation of the refined solution, and calculation of the RI subject to available measurements continues till iteration~7. At iteration~7 the yellow RoI uses the last measurements to calculate the RI for a refinement to a macro pixel resolution of~$1 \times 1$. At iteration~8 the orange RoI is resolved to~$1 \times 1$ macro pixel resolution as the required measurements were already acquired at iteration~5. At iteration~9 the green RoI is resolved to a macro pixel resolution of~$2 \times 2$ but further refinement is not possible as the measurement budget has been exhausted. Therefore, the final macro pixel resolution of the green RoI is~$2 \times 2$. Finally, the yellow RoI is resolved to the native resolution of the DMD by using the refinement measurements collected at iteration~7. This shows the dynamic distribution of the measurements across the RoIs by the RPS algorithm. For acquisition of all the RoIs at the native resolution of the DMD, one would need~10000 measurements. Thus, the RPS algorithm selects the most informative RoI at each iteration and better utilises the limited measurement budget (8060 measurements). 

Due to space limitations the application of RPS on the Gulf Vis dataset is shown in Fig.~\ref{figGulfVisRI_rois} and Table~\ref{tableGulfVisRIValues} in the supplementary material. The results for Pavia are provided in Fig.~\ref{figPaviaRI_rois} and Table~\ref{tablePaviaRIValues} in the supplementary material. In the case of Pavia we can see that a smaller RoI (RoI 1 and RoI 4) with more information change across the spatial scales is prioritised over a larger sized RoI (RoI 3).

\begin{table}
\caption[Evolution of RI for Cuprite IR dataset.]{Evolution of RI for Cuprite IR dataset in Fig.~\ref{figCupIRRI_limitedBudget}. The highlighted RoI indicates the one selected at each iteration. The refined macro pixel resolution of the selected RoI is tabulated. The total number of measurements is 8060.}
\begin{center}
\newcolumntype{P}[1]{>{\centering\arraybackslash}p{#1}}
%\begin{tabularx}{\columnwidth}{|>{\centering}X|>{\centering}X|>{\centering}X|>{\centering}X|>{\centering}X|>{\centering\arraybackslash}X|}
\begin{tabularx}{\columnwidth}{|>{\centering}X|>{\centering}X|>{\centering}X|>{\centering}X|>{\centering}X|>{\centering}X|>{\centering\arraybackslash}p{1cm}|}
\hline
	& \multicolumn{4}{c|}{RI} & &\\	
%\center{Iteration No.}& \multicolumn{4}{c|}{\thead{RI}} &Refined Macro Pixel Resolution & Available Measurements\\	
	%\hline
	\cline{2-5}
%& RoI 1 (Red) & RoI 2 (Orange) & RoI 3 (Green) & RoI 4 (Yellow)&  &\\
%\center{Iteration No.}& \thead{RoI 1\\ ( Red )} & \thead{RoI 2 \\ (Orange)} &\thead{ RoI 3 \\ (Green)} & \thead{ RoI 4 \\(Yellow)}&Refined Macro Pixel Resolution &Available Measurements\\
%\makecell{\hspace{-1.5mm} Iteration\\ \hspace{-1.5mm} No.}& \! \makecell[{{m{1cm}}}]{\ RoI 1\\ \ (Red)\\ \hspace{-2.2mm}(\(128 \! \times\! 128\))} &\! \makecell[{{m{1cm}}}]{\ RoI 2\\ \hspace{-1mm}(Orange)\\\hspace{-1.6mm}($64\, \times \,64$)} & \! \makecell[{{m{1cm}}}]{\ RoI 3\\\hspace{-0.2mm}(Green)\\\hspace{-1.8mm}($32\, \times\, 32$)} & \! \makecell[{{m{1cm}}}]{\ RoI 4\\ \hspace{-0.5mm}(Yellow)\\ \hspace{-1.7mm}($32\, \times\, 32$)}&\vspace{-1mm} Refined Macro Pixel Resolution \vspace{1mm} & \vspace{-1mm}Available Measurements\\
\makecell{\hspace{-1.7mm} Iteration \\No.}& \! \makecell[{{m{1cm}}}]{\ RoI 1\\ \ (Red)\\ \hspace{-2.2mm}(\(128 \! \times\! 128\))} &\! \makecell[{{m{1cm}}}]{\ RoI 2\\ \hspace{-1mm}(Orange)\\\hspace{-1.6mm}($64\, \times \,64$)} & \! \makecell[{{m{1cm}}}]{\ RoI 3\\\hspace{-0.2mm}(Green)\\\hspace{-1.8mm}($32\, \times\, 32$)} & \! \makecell[{{m{1cm}}}]{\ RoI 4\\ \hspace{-0.5mm}(Yellow)\\ \hspace{-1.7mm}($32\, \times\, 32$)}&\vspace{-1mm} Refined Macro Pixel Resolution \vspace{1mm} & \vspace{-1mm}Available Measurements\\
\hline

1 & \cellcolor{green} 272.21 & 139.14 &8.39 & 29.55  & $4 \times 4$&2762\\
\hline
%2 & \cellcolor{green} $3.5 \times 10^4$ & $2.5 \times 10^4$& $1.5 \times 10^3$& 807.9 & $2\times 2$&920\\
2 &  41.43 & \cellcolor{green} 139.14& 8.39& 29.55 & $4\times 4$&1124\\
\hline
%3 & - &  \cellcolor{green}$2.5 \times 10^4$ & $1.5 \times 10^3$&807.9 &$4 \times 4$&920 \\
3 & \cellcolor{green}41.43 & 14.04 & 8.39&29.55 &$2 \times 2$&715 \\
\hline
%4 &  -& \cellcolor{green}$3 \times 10^3$ &   $1.5 \times 10^3$&807.9 &$2\times 2$ &511\\
4 &  -&14.04 &8.39&\cellcolor{green}29.55 &$4\times 4$ &715\\
\hline
%5 & - & 619.9 &  \cellcolor{green}$1.5 \times 10^3$& 807.9 & $4 \times 4$&102 \\
5 & - &\cellcolor{green} 14.04 & 8.39&4.13 & $2 \times 2$&613 \\
\hline
6 & - & 3.44 &\cellcolor{green}8.39&4.13&  $4 \times 4$&204 \\
\hline
7 & - & 3.44& 1.80&\cellcolor{green}4.13 & $2 \times 2$&102 \\
\hline
8 &- &\cellcolor{green}3.44&1.80 &1.10&$1 \times 1$&0\\
\hline
9& - & - &\cellcolor{green}1.80 &1.10&$2 \times 2$&0\\
\hline
10& - & - &- &\cellcolor{green}1.10&$1 \times 1$&0\\
\hline
11& - & - &- &-&-&0\\
\hline
\end{tabularx}

\label{tableCupIRRI_limitedMeasurement}
\end{center}
\end{table}

\begin{figure}[!t]
\centering
\includegraphics[width=\columnwidth, height = 10cm, keepaspectratio]{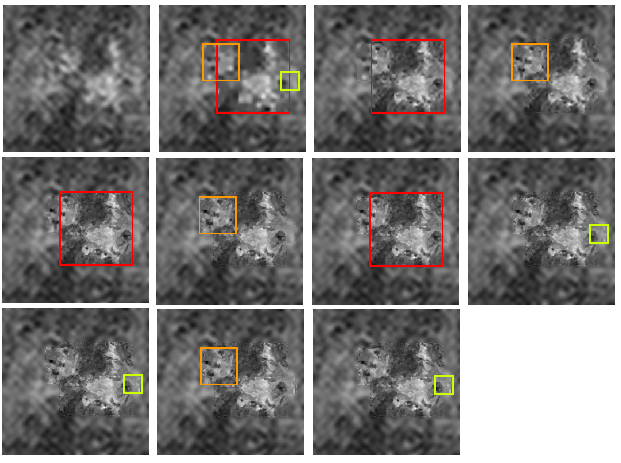}
\caption[Prioritisation of RoIs with RI for Cuprite Vis dataset with Rademacher measurements.]{Prioritisation of RoIs with RI for Cuprite Vis dataset with random-macro-pixel Rademacher measurements. The total number of measurements is 9188 (14\% of the total number of pixels). The number of physical measurements is 18376 (28\% of the total number of pixels).}
\label{figCupVisRI_Rademacher}
\end{figure}

% Rademacher n walsh
\subsection{Experiments with Different Measurement Matrices}
\label{measMatrixExp}

The RPS is agnostic to the nature of the measurement matrix. The only requirement is that the measurement ensemble can be deployed in multi-scale manner. Fig.~\ref{figCupVisRI_Rademacher} shows the result of application of the RPS on the Cuprite Vis dataset with Rademacher ($\pm1$ with equal probability)~\cite{Foucart} random-macro-pixel measurements. The RoIs and their order of prioritisation changes, though the overall result is similar to $0/1$ random-macro-pixel sampling. The corresponding error metrics are shown in Table~\ref{tableCupVisRI_Rademacher}. The first step of low resolution acquisition for detecting RoIs is still performed with $0/1$ random-macro-pixel sampling. Implementation of Rademacher random-macro-pixel measurements requires two acquisition cycles on the DMD due to the $\pm1$ nature of the Rademacher random variable. The total number of physical measurements required for Rademacher measurements is twice the number of realisations of the measurement vectors. %Let random-macro-pixel sampling require $p\%$ of the number of pixels in the selected RoI for calculation of RI at each spatial resolution.
Let the number of measurements required for calculation of RI at each spatial scale for the $i^{th}$~RoI be $p_i$\%  of the total number of pixels in the RoI. Then, for $M$~RoIs %with $p_i$\% measurements needed for each RoI per spatial resolution%
 with $R$ resolution levels, the total number of $0/1$ random-macro-pixel measurements will be ${CM + R\sum_{i=1}^M \left\lfloor \frac{p_i}{100} n_i \right\rfloor }$, where $CM$ is the number of low resolution measurements common for all RoIs and $n_i$ denotes the total number of pixels in each RoI. For Rademacher random-macro-pixel measurements, the total number of physical measurements, i.e., the number of acquisition cycles performed on the DMD, will be equal to $ CM + 2R \sum_{i=1}^M \left\lfloor \frac{p_i}{100} n_i \right\rfloor$.

\begin{table}
\caption{Error metrics for the RoIs for Cuprite Vis dataset with random-macro-pixel Rademacher measurements (Fig.~\ref{figCupVisRI_Rademacher}). Resolution is in macro pixel size.}
\begin{center}
\newcolumntype{P}[1]{>{\centering\arraybackslash}p{#1}}
\begin{tabularx}{\columnwidth}{|P{1.1cm}|>{\centering}X|>{\centering}X|>{\centering}X|>{\centering}X|>{\centering}X|>{\centering\arraybackslash}X|}
\hline
Resolution of RoI	& \multicolumn{2}{>{\centering\setlength\hsize{2\hsize} }X|}{\thead{RoI 1 \\(Red)\\($128 \times 128$)}}& \multicolumn{2}{>{\centering\setlength\hsize{2\hsize} }X|}{\thead{RoI 2 \\ (Orange) \\ ($64 \times 64$)}}& \multicolumn{2}{>{\centering\setlength\hsize{2\hsize} }X|}{\thead{RoI 3 \\(Yellow)\\ ($32 \times 32$)}}\\	
%	\hline
\cline{2-7}
  & NMSE & SSIM & NMSE & SSIM & NMSE & SSIM  \\
	\hline
$8 \times 8$& 0.032 &0.39 &0.043&0.35 & 0.022& 0.4 \\
\hline
$4 \times 4$&0.015&0.63 &0.017&0.63 &0.015 &0.54\\
\hline
$2 \times 2$&0.008 & 0.78& 0.009&0.79 & 0.007& 0.77\\
\hline
$1 \times 1$& 0.006& 0.83 & 0.007& 0.83 &0.005 &0.83 \\
\hline
\end{tabularx}

\label{tableCupVisRI_Rademacher}
\end{center}
\end{table}

\begin{figure}[!t]
\centering
\includegraphics[width=\columnwidth, keepaspectratio]{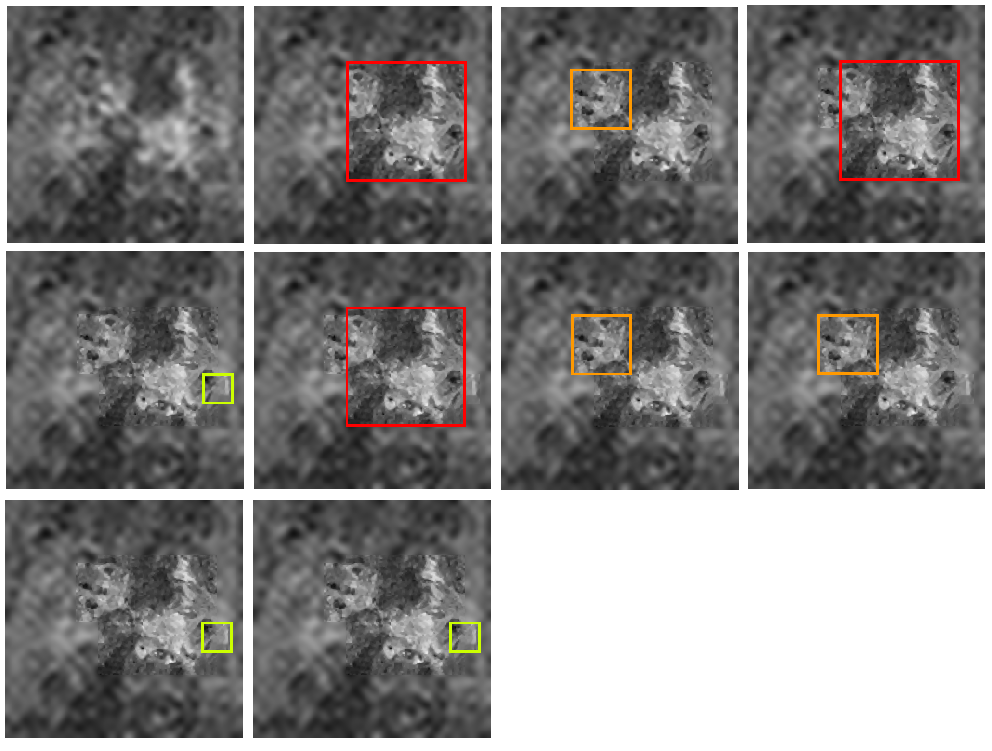}
\caption[Prioritisation of RoIs with RI for Cuprite Vis dataset with Walsh measurements exploiting multi-level sampling.]{Prioritisation of RoIs with RI for Cuprite Vis dataset with Walsh measurements exploiting multi-level sampling. The total number of measurements is 5300 (8.1\% of the total number of pixels). The number of physical measurements is 9600 (14.6\% of the total number of pixels).}
\label{figCupVisRI_walsh}
\end{figure}

One can also use Walsh transforms in a multi-scale manner. The Walsh sampling masks from Fig.~\ref{fig:samplingMaps} are used. The number of Walsh measurements in each spatial frequency band is provided in Table~\ref{tableWalshMeasurementsPerBand}. For each RoI size the total number of Walsh measurements is fixed beforehand while determining the sampling map. The number of measurements required for RPS with Walsh ensemble is equal to $CM + \sum_{i=1}^M \left\lfloor \frac{p_i}{100}n_i \right\rfloor $. The total physical measurements cycles on the DMD is twice the second term, as the Walsh ensemble consists of $\pm1$ measurements, plus the low resolution measurements for the entire scene. The low resolution acquisition is performed with $0/1$ random-macro-pixel measurements. For RoI reconstruction only the Walsh measurements are used. For calculating the initial RI values, the low frequency Walsh measurements are used as the refined measurements and the low resolution regions corresponding to the RoIs from the low resolution reconstruction are used as the coarse reconstruction for the \textit{simulated} measurements. For further refinement the Walsh reconstructions at different spatial scales are used as the coarse reconstructions for the \textit{simulated} measurements. The results with multi-level Walsh measurements are shown in Fig.~\ref{figCupVisRI_walsh} and in Table~\ref{tableCupVisRI_walsh}. The reconstruction metrics are recorded in Table~\ref{tableCupVisRI_roiMetricWalsh}. The advantage of using the concept of multi-level sampling with the concept of RI is evident from comparing the results in Table~\ref{tableCupVisRI_Rademacher} and Table~\ref{tableCupVisRI_roiMetricWalsh}. Multi-level Walsh measurements with RI achieve similar performance as random-macro-pixel Rademacher measurements with a smaller measurement budget.

\begin{table}
\caption[Number of measurements per spatial frequency band in Walsh sampling masks.]{Number of measurements per spatial frequency band in Walsh sampling masks.}
\begin{center}
\begin{tabularx}{\columnwidth}{|>{\centering}X|>{\centering}X|>{\centering}X|>{\centering\arraybackslash}X|}
\cline{2-4}
	\multicolumn{1}{c}{}& \multicolumn{3}{|c|}{Number of Required Measurements} \\	
	\hline
Mask Size & Low Frequency & Medium Frequency & High Frequency  \\
\hline
$128 \times 128$& 2025 & 689 & 562\\
\hline
$64 \times 64$& 676 & 79 & 64\\
\hline
$32 \times 32$& 121 & 67 & 16\\
\hline
\end{tabularx}

\label{tableWalshMeasurementsPerBand}
\end{center}
\end{table}

\begin{table}
\caption[Evolution of RI for Cuprite Vis dataset with Walsh measurements using multi-level sampling.]{Evolution of RI for Cuprite Vis dataset with Walsh measurements using multi-level sampling (Fig.~\ref{figCupVisRI_walsh}). The highlighted RoI indicates the one selected at each iteration.}
\begin{center}
\begin{tabularx}{\columnwidth}{|>{\centering}p{0.8cm}|>{\centering}X|>{\centering}X|>{\centering}X|>{\centering\arraybackslash}X|}
\hline
%& \multicolumn{3}{c|}{RI} &\\	
\thead{Iteration \\  No.}& \multicolumn{3}{c|}{\thead{RI}} &\thead{Available\\Measurements}\\	
%	\hline
\cline{2-4}
%\center{Iteration No.}& \thead{RoI 1 \\ (Red)\\ ($128 \times 128$)} & \thead{ RoI 2 \\ (Orange) \\ ($64  \ \times\ 64$)} & \thead{RoI 3\\ (Yellow)\\ ($32 \ \times \ 32$) }& Available Measurements \\
& \thead{RoI 1 \\ (Red)\\ ($128 \times 128$)} & \thead{ RoI 2 \\ (Orange) \\ ($64  \ \times\ 64$)} & \thead{RoI 3\\ (Yellow)\\ ($32 \ \times \ 32$) }& \\

	\hline
1 & \cellcolor{green} 520.56 & 110.12 & 24.83& 1478 \\
	\hline
2 &  81.84 & \cellcolor{green}110.12& 24.83& 789 \\
\hline
3 & \cellcolor{green}81.84 &  18.15 & 24.83&710 \\
\hline
4 &  22.44 &  18.15  & \cellcolor{green} 24.83&148 \\
\hline
5 & \cellcolor{green} 22.44&  18.15  & 5.50 &81\\
\hline
6 & - & \cellcolor{green}18.15&5.50&81 \\
\hline
7 & - & \cellcolor{green} 7.30 & 5.50&17 \\
\hline
8 & - & - & \cellcolor{green}5.50&17 \\
\hline
9 & - & - & \cellcolor{green}3.54 &1\\
\hline
10 & -  & - & - & 1 \\
\hline
\end{tabularx}

\label{tableCupVisRI_walsh}
\end{center}
\end{table}

\begin{table}
\caption{Error metrics for the RoIs for Cuprite Vis dataset with Walsh measurements (Fig.~\ref{figCupVisRI_walsh}) using multi-level sampling in the RPS algorithm.}
\begin{center}
\newcolumntype{P}[1]{>{\centering\arraybackslash}p{#1}}
%\begin{tabularx}{\columnwidth}{|P{3.5cm}|P{1.1cm}|P{1.1cm}|>{\centering}X|>{\centering}X|>{\centering}X|>{\centering\arraybackslash}X|}
\begin{tabularx}{\columnwidth}{P{1.1cm}|>{\centering}X|>{\centering}X|>{\centering}X|>{\centering}X|>{\centering}X|>{\centering\arraybackslash}X|}
%\begin{tabularx}{\columnwidth}{P{1.1cm}|P{1cm}|>{\centering}X|>{\centering}X|>{\centering}X|>{\centering}X|>{\centering\arraybackslash}X|}
\cline{2-7}
%\hline \centering
	& \multicolumn{2}{>{\centering\setlength\hsize{2\hsize} }X|}{\thead{RoI 1\\(Red)\\ ($128  \times  128$)}}& \multicolumn{2}{>{\centering\setlength\hsize{2\hsize} }X|}{\thead{RoI 2 \\(Orange)\\ ($64 \ \times \ 64$)}}& \multicolumn{2}{>{\centering\setlength\hsize{2\hsize} }X|}{\thead{RoI 3\\ (Yellow)\\ ($32 \ \times \ 32$)}}\\	
	\cline{2-7}
 & NMSE & SSIM & NMSE & SSIM & NMSE & SSIM  \\
	\hline
\multicolumn{1}{|c|}{Coarse Step}& 0.031 &0.42 &0.031&0.45 & 0.022& 0.41 \\
\hline
\multicolumn{1}{|c|}{Refinement Step 1} &0.008 &0.79 &0.007&0.82 &0.009  &0.71 \\
\hline
\multicolumn{1}{|c|}{Refinement Step 2}&0.007 & 0.82& 0.007&0.83 & 0.007& 0.76\\
\hline
\multicolumn{1}{|c|}{Refinement Step 3}& 0.006& 0.83 & 0.007& 0.84 &0.006 &0.77\\
\hline
\end{tabularx}
\label{tableCupVisRI_roiMetricWalsh}
\end{center}
\end{table}

\subsection{Results on MSL Images}

\begin{figure}[!t]
\centering
\includegraphics[width=\columnwidth, keepaspectratio]{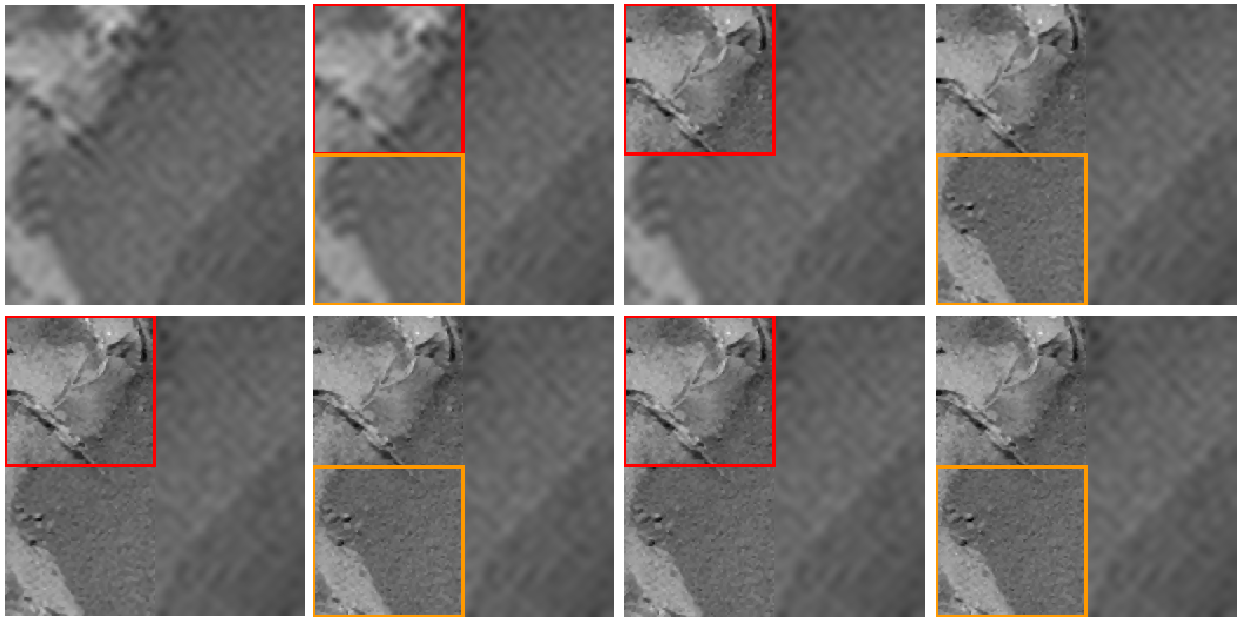}
\caption[Prioritisation of RoIs with RI for Mars1 dataset.]{Prioritisation of RoIs with RI for Mars1 image. The total number of measurements is 7552 (11.5\% of the total number of pixels). }
\label{fig:riEvolution_mars1}
\end{figure}

\begin{figure}[!t]
\centering
\includegraphics[width=\columnwidth, keepaspectratio]{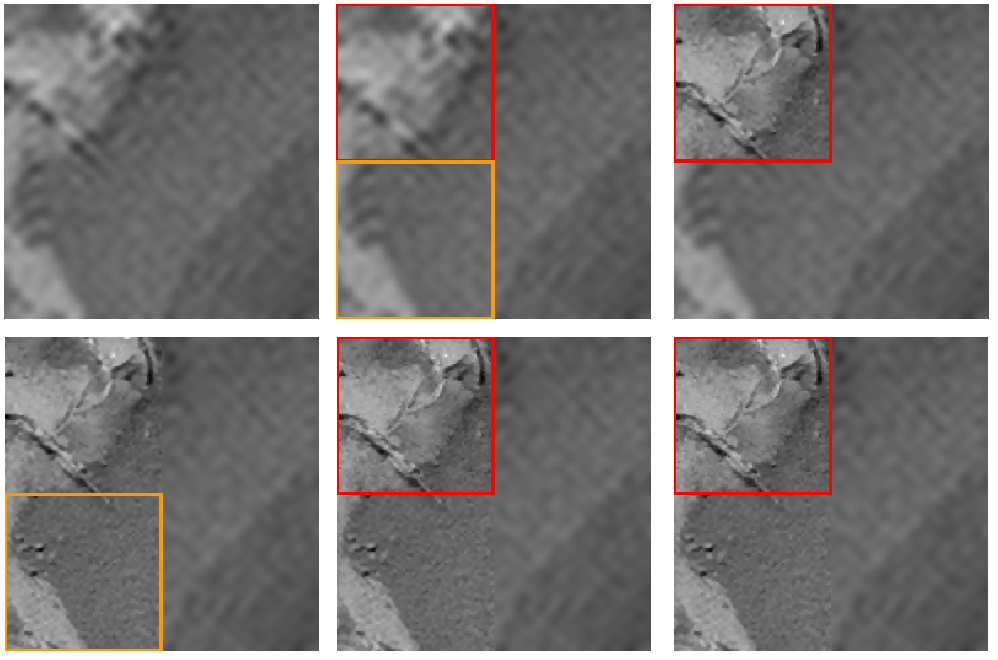}
\caption[Prioritisation of RoIs with RI for Mars1 dataset with limited budget.]{Prioritisation of RoIs with RI for Mars1 image with limited measurement budget. The total number of measurements is~6301 (9.6\% of the total number of pixels). }
\label{fig:riEvolution_mars1_limited}
\end{figure}

\begin{figure}[!t]
\centering
\includegraphics[width=\columnwidth, keepaspectratio]{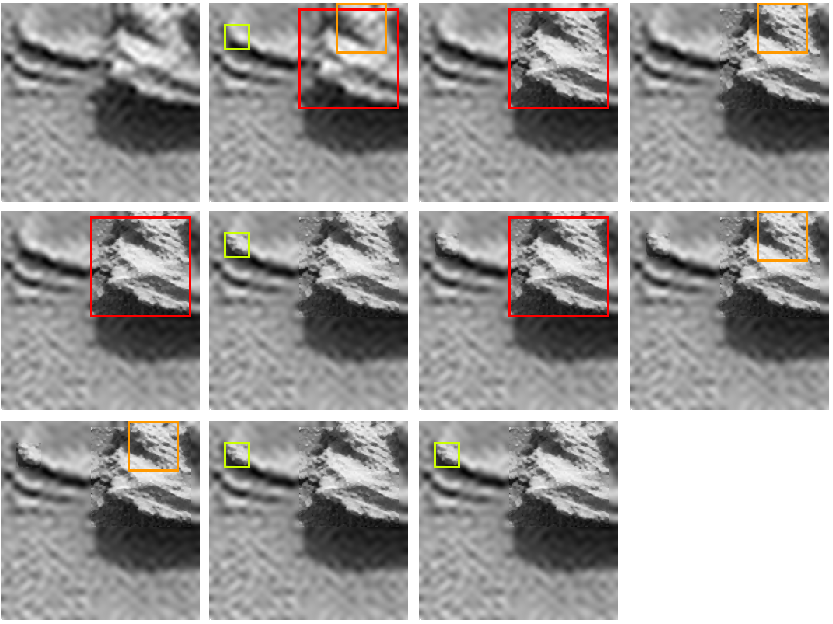}
\caption[Prioritisation of RoIs with RI for Mars2 dataset.]{Prioritisation of RoIs with RI for Mars2 image. The total number of measurements is 5299 (8.1\% of the total number of pixels). }
\label{fig:riEvolution_mars2}
\end{figure}

This section reports the results of using RPS to reconstruct the Mars1 and Mars2 images from Walsh measurements. The prioritisation of RoIs for Mars1 image is shown in Fig.~\ref{fig:riEvolution_mars1}. The evolution of RI values and RoI prioritisation are recorded in Table~\ref{table:riEvolution_mars1}. A limited budget experiment is also conducted with the Mars1 image where the number of measurements is not enough to resolve each RoI to the native resolution of the DMD. Fig.~\ref{fig:riEvolution_mars1_limited} and Table~\ref{table:riEvolution_mars1_limited} show the prioritisation of the RoIs and the evolution of the RI values respectively. The orange RoI is resolved using only the low frequency Walsh measurements denoted by the green region in Fig.~\ref{fig:samplingMaps}(c) and the corresponding number of measurements is given in Table~\ref{tableWalshMeasurementsPerBand}. As there is not enough measurements for further refinement of the orange RoI, it is not resolved using the higher frequency Walsh measurements and is removed from the list. Only the red RoI is resolved to the native resolution of the DMD. Thus, the RI is able to prioritise RoIs on the basis of the information change across spatial scales of the RoIs from Mars rover camera images as well. Error metrics for both measurement budget scenarios of the Mars1 image are provided in Table~\ref{tableMars1RI_roiMetricWalsh} and Table~\ref{tableMars1RI_roiMetricWalsh_limited} in the supplementary material. The result of applying the RPS on the Mars2 image is shown in Fig.~\ref{fig:riEvolution_mars2}. The evolution of the RI and RoI selection is shown in Table~\ref{tableRIEvolution_Mars2} in the supplementary material. The error metrics are recorded in Table~\ref{tableMars2RI_roiMetricWalsh} in the supplementary material.

\begin{table}
\caption[Evolution of RI for Mars1 dataset.]{Evolution of RI for Mars1 image with Walsh measurements using multi-level sampling (Fig.~\ref{fig:riEvolution_mars1}). The highlighted RoI indicates the one selected at each iteration. The total number of measurements is 7552.}
\begin{center}
\begin{tabularx}{\columnwidth}{|>{\centering}X|>{\centering}X|>{\centering}X|>{\centering\arraybackslash}X|}
\hline
%	& \multicolumn{2}{c|}{RI} &\\	
		\thead{Iteration \\No.} & \multicolumn{2}{c|}{\thead{RI}} & \thead{Available \\Measurements} \\
%\hline
\cline{2-3}
%Iteration No.&\thead{ RoI 1 \\(Red)\\ ($128 \times 128$)} & \thead{RoI 2\\ (Orange)\\ ($128 \times 128$)} & Available Measurements  \\
&\thead{ RoI 1 \\(Red)\\ ($128 \times 128$)} & \thead{RoI 2\\ (Orange)\\ ($128 \times 128$)} &   \\
	\hline
1 & \cellcolor{green} 293.72 & 147.39  & 2502\\
\hline
2 &  64.84 & \cellcolor{green} 147.39&1813\\
\hline
3 & \cellcolor{green}64.84 & 49.33 & 1124 \\
\hline
4 &  14.37 & \cellcolor{green}49.33 &562\\
\hline
5 & \cellcolor{green}14.37 & 13.75 & 562 \\
\hline
6 & - & \cellcolor{green}13.75 &  0 \\
\hline
7 & - & -  & 0\\
\hline
\end{tabularx}

\label{table:riEvolution_mars1}
\end{center}
\end{table}

\begin{table}
\caption[Evolution of RI for Mars1 dataset with limited measurements.]{Evolution of RI for Mars1 image with Walsh measurements using multi-level sampling (Fig.~\ref{fig:riEvolution_mars1_limited}) with limited measurement budget. The highlighted RoI indicates the one selected at each iteration. The total number of measurements is 6301.}
\begin{center}
\begin{tabularx}{\columnwidth}{|>{\centering}X|>{\centering}X|>{\centering}X|>{\centering\arraybackslash}X|}
\hline
%	& \multicolumn{2}{c|}{RI} &\\	
	\thead{Iteration \\No.} & \multicolumn{2}{c|}{\thead{RI}} & \thead{Available \\Measurements} \\	
%\hline
\cline{2-3}
%Iteration No.& \thead{ RoI 1 \\ (Red)\\ ($128 \times 128$)} & \thead{RoI 2 \\(Orange) \\($128 \times 128$)} & Available Measurements  \\
& \thead{ RoI 1 \\ (Red)\\ ($128 \times 128$)} & \thead{RoI 2 \\(Orange) \\($128 \times 128$)} &   \\

	\hline
1 & \cellcolor{green} 293.72 & 147.39  & 1251\\
\hline
2 &  64.84 & \cellcolor{green} 147.39 & 562 \\
\hline
3 & \cellcolor{green}64.84 & - & 562 \\
\hline
4 &  \cellcolor{green}14.37 & - &562\\
\hline
5 & - & - & 0 \\
\hline
\end{tabularx}

\label{table:riEvolution_mars1_limited}
\end{center}
\end{table}

\subsection{Discussion: Noisy Measurements and Processing Time}
\label{sec:Noise_time}

In case of noisy observations, the RPS is still applicable. The error estimate can be used in the same way, since the RI only takes into account the given data independent of their model. The reason is that we approximate regularised solutions at finer levels (that is an estimator based on the given noisy data) instead of an idealised solution. Consequently, the output of the ``detection and segmentation" step may lead to different RoI selections with the noise realisations, but we expect the first ones to remain more stable. Deviations from an idealised solution are rather a model error than due to our adaptive algorithm, the issue of simultaneously treating modelling errors is beyond the scope of the paper and left to future research. Examples of RPS with noisy observations using the Cuprite~Vis dataset are provided in Fig.~\ref{figCupVisNoise1},~\ref{figCupVisNoise2},~\ref{figCupVisNoise3}, and~\ref{figCupVisNoise4} in the supplementary material. In case of low light conditions, if the photon count is not too low then the noise can be modelled as a Gaussian with a variable variance and the RI is applicable. Analysing scenarios dominated by Poisson noise is beyond the scope of the article, it will be considered in future work along the lines of~\cite{errEstimatePoisson}.

The required processing time depends on the underlying compressed sensing solver. This can be tailored to the required application and may be iterative or learned in nature. If the application admits structured measurement matrices, such as the Walsh or Fourier matrices, then the FFT algorithm can be exploited to accelerate the reconstruction. The complexity analysis of compressed sensing solvers can be found in~\cite{tfocs} and the references therein.  Calculation of the RI involves a vector subtraction that takes $N$ FLOPs and a calculation of a squared $l_2$ norm that takes $2N-1$ FLOPs, where $N$ is the total number of pixels in the RoI. The total number of FLOPs required for RI calculation is $3N-1$.

\subsection{Comparison with Other Techniques}

As discussed previously, for exploration scenarios region based acquisition is preferred over acquisition of the complete scene due to limited resources. Also a region based approach provides the opportunity to acquire the RoIs with greater precision. A comparison of the RPS with classical compressed sensing~\cite{CRT1,donoho1} and multi-level compressed sensing~\cite{asymptoticSparsity} with Walsh measurements is done in Table~\ref{tableCompCupVisCs}. The measurement budget is 5300 for all three acquisition methods and algorithm~(\ref{reviewEq15}) is used with Daubechies-8 wavelets as the sparsity basis.

Rademacher measurements are used for classical compressed sensing.  On its own classical compressed sensing and multi-level compressed sensing recover the entire scene at the native resolution of the DMD. As expected the RoIs are best reconstructed by the RPS, as in that case the RoIs are recovered in a dedicated manner. As the background is left at a lower resolution in the RPS algorithm the recovery of the background is best with multi-level compressed sensing. Table~\ref{tableCompCupVisCs} shows that in an exploration scenario with limited measurement budget, region based acquisition, like the RPS, is able to distribute the measurement budget intelligently and acquire better quality RoIs, as measured by NMSE and SSIM. Fig.~\ref{figCompCupVisCS} in the supplementary material shows the reconstruction comparison from the three considered methods. 

\begin{table}
\caption{Error metrics for comparison of RPS with classical and multi-level Compressed Sensing (CS).}
\begin{center}
\newcolumntype{P}[1]{>{\centering}p{#1}}
\begin{tabularx}{\columnwidth}{>{\centering}X >{\centering}X|>{\centering}X|>{\centering}X|>{\centering\arraybackslash}X|}
\cline{3-5}
& & Classical CS & Multi-level CS & RPS \\
\cline{1-5}
\multicolumn{1}{|c|}{\multirow{4}{1cm}{NMSE} }& 
{RoI 1 (Red)}& 0.016 & 0.011 & 0.006 \\ \cline{2-5}
\multicolumn{1}{|c|}{}&{RoI 2 (Orange)} & 0.017 & 0.011 & 0.007 \\ \cline{2-5}
\multicolumn{1}{|c|}{}&{RoI 3 (Yellow)} & 0.016 & 0.011 & 0.006 \\ \cline{2-5}
\multicolumn{1}{|c|}{}&{Background} & 0.029 & 0.012 & 0.028 \\ \cline{1-5}

\multicolumn{1}{|c|}{\multirow{4}{1cm}{SSIM} }& 
{RoI 1 (Red)}& 0.61 & 0.71 & 0.83 \\ \cline{2-5}
\multicolumn{1}{|c|}{}&{RoI 2 (Orange)} & 0.65 & 0.73 & 0.84 \\ \cline{2-5}
\multicolumn{1}{|c|}{}&{RoI 3 (Yellow)} & 0.65 & 0.71 & 0.77 \\ \cline{2-5}
\multicolumn{1}{|c|}{}&{Background} & 0.7 & 0.79 & 0.62 \\ \cline{1-5}

\end{tabularx}

\label{tableCompCupVisCs}
\end{center}
\end{table}

\begin{figure}[!t]
\centering
\includegraphics[width=\columnwidth, keepaspectratio]{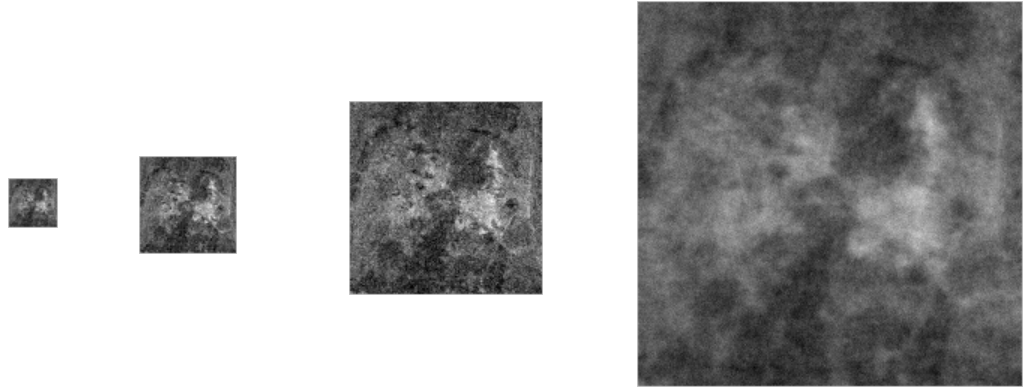}
\caption[MR-AMP for Cuprite Vis.]{MR-AMP reconstruction for Cuprite Vis dataset. From left to right down-sampling factor: 8, 4, 2, and 1 . The total number of measurements is 9600 (14.6\% of the total number of pixels).}
\label{figMRAMP}
\end{figure}
\begin{figure}[!t]
\centering
\includegraphics[width=\columnwidth, keepaspectratio]{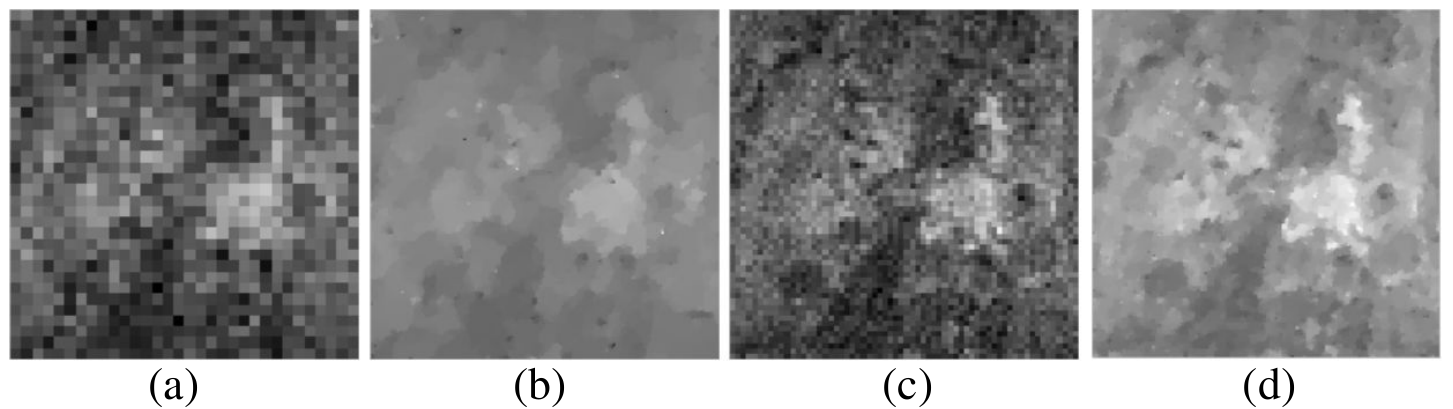}
\caption[Stone for Cuprite Vis.]{STOne transformation reconstruction for Cuprite Vis dataset. (a) Low resolution reconstruction with 1024 measurements, (b) high resolution reconstruction of (a), low resolution reconstruction with 4096 measurements, and (d) high resolution reconstruction of (c).}
\label{figStone}
\end{figure}

Fig.~\ref{figMRAMP} and Fig.~\ref{figStone} show the multi-resolution compressed sensing reconstructions from MR-AMP~\cite{ampMR, mrampCode} and the STOne transform~\cite{stoneTransform, stoneCode} respectively for the Cuprite Vis dataset.  MR-AMP used the same number of measurements for reconstruction at different down-sampling factors as the RPS algorithm illustrated in Fig.~\ref{figCupVisRI}. For STOne transformation the number of measurements must be a power of 2, therefore, we use 1024 and 4096 measurements to reconstruct two different low resolution images. These comparisons show that at the considered sub-sampling rates, a step-wise resolution refinement procedure, like the RPS, works better. No codes were provided by the authors of ~\cite{multiResCompSensing}, therefore, no comparison was possible. 

\section{Conclusion}

This work proposes a procedure for RoI prioritisation in constrained systems. An estimator for the change in information content across spatial resolutions is developed. The Refinement Indicator (RI) evaluates the importance of a RoI for refinement to a finer resolution by calculating the error between the novel refined measurements and \textit{simulated} high resolution measurements obtained by considering the coarse resolution reconstruction as the groundtruth. Thus, the calculation of the refined higher spatial resolution reconstruction is not required to estimate the change in information content across scales. The RI is used to prioritise RoIs for sequential acquisition through the RPS algorithm. Experiments show that the RPS algorithm is able to distribute the limited measurement budget in constrained systems in an organized manner across the different RoIs. At the considered sampling rates, the RoI based step-wise resolution refinement performs better at acquiring mission relevant RoIs than the state-of-the-art multi-resolution reconstruction techniques that recover the complete scene at a low resolution. Multi-scale multi-level compressed sensing with Walsh measurements is employed in the RPS framework and provides similar performance to random-macro-pixel sampling using Rademacher measurements with smaller measurement budgets. The RI thus provides a method for dynamic measurement budget allocation in constrained systems, which increases the capabilities of autonomous systems in exploration scenarios.

In the future, the work may be extended to multispectral scenarios and to low photon count environments. Methods for sequential compressed sensing reconstruction~\cite{regmodcs} may be studied for this application.

\bibliographystyle{IEEEtran}
\bibliography{IEEEabrv,tci_bib_rpmcs}
%\vspace{1cm}
\begin{IEEEbiography}[{\includegraphics[width=1in,height=1.25in,clip,keepaspectratio]{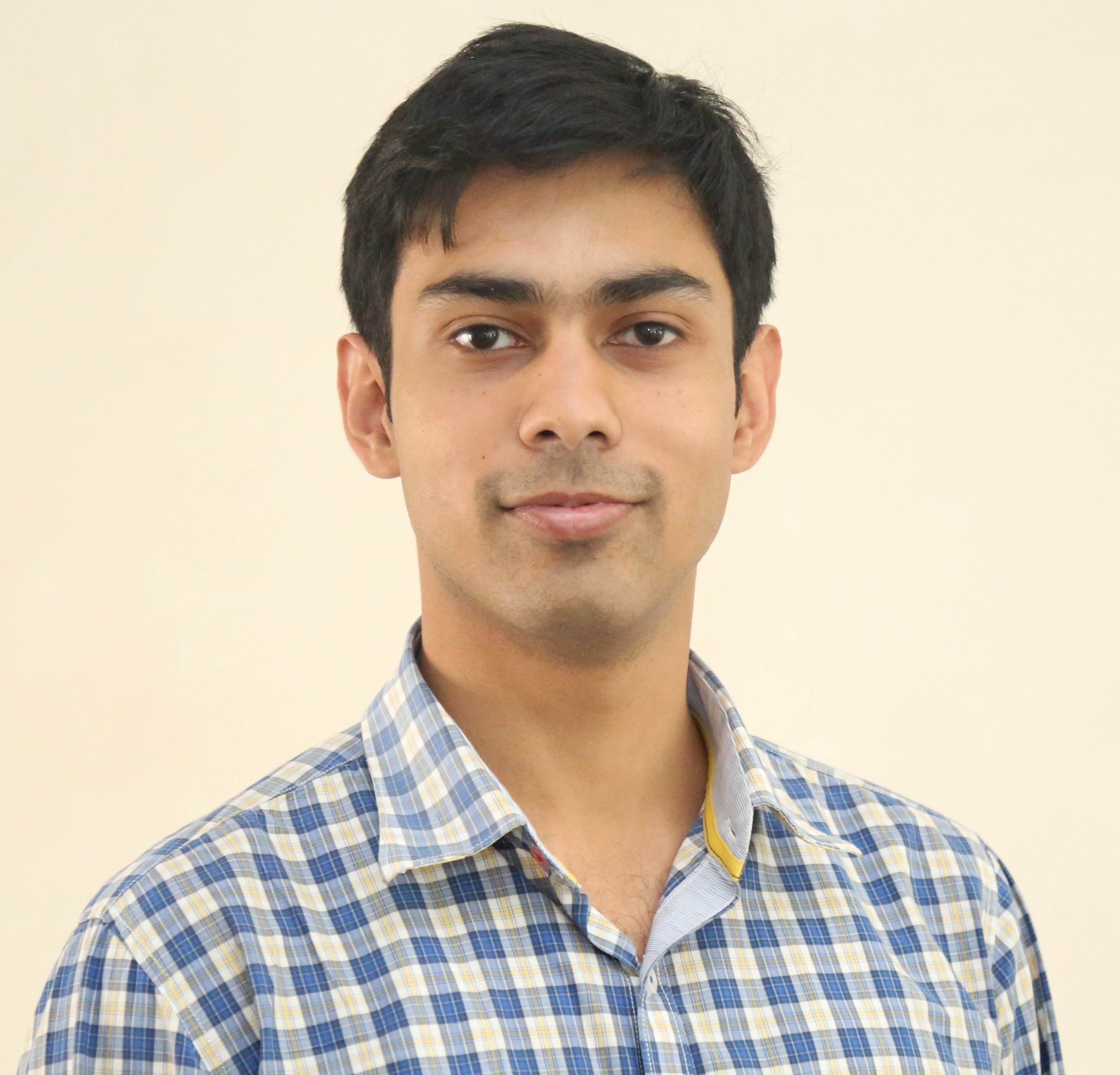}}]{Protim Bhattacharjee} was born in New Delhi, India, in 1991. He studied electronics and communications engineering at IIIT Delhi, New Delhi, India. 
He currently works at the German Aerospace Center (DLR), Berlin, Germany and is a doctoral student at the department of electrical engineering at the Friedrich-Alexander-University Erlangen-Nürnberg (FAU). His research interests include computational imaging, statistical estimation and signal processing methods for remote sensing.
\end{IEEEbiography}
%\vspace{-1.2cm}
\begin{IEEEbiography}[{\includegraphics[width=1in,height=1.25in,clip,keepaspectratio]{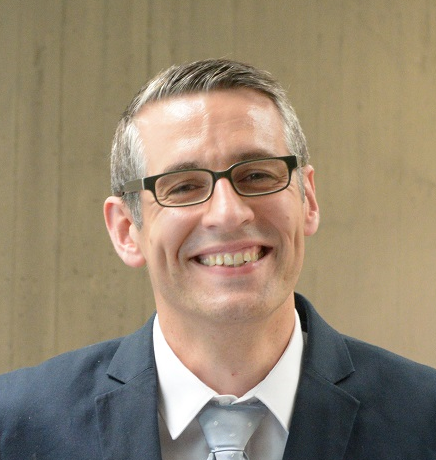}}]{Martin Burger} was born in Wels, Austria, in 1976. He studied mathematics at Johannes Kepler University Linz and University of Milano, and finished his PhD in 2000. 
After working as a postdoc at Johannes Kepler University, CAM Assistant Professor at UCLA, and scientific adviser of the Johann Radon Institute for Computational and Applied Mathematics, he accepted an offer from the Westfälische-Wilhelms University in Münster, Germany, where he worked as full professor from 2006 to 2018. Since 2018 he is full professor in applied mathematics at Friedrich-Alexander-University Erlangen-Nürnberg (FAU). His research interests include inverse problems, mathematical imaging,  and theoretical foundations of deep learning. 
Dr. Burger received numerous awards and honours for his work in applied mathematics and inverse problems such as the Calderon Prize of the International Problems International Association and an ERC Consolidator Grant in 2013. He serves in the editorial board of several journals in applied mathematics including Inverse Problems, Inverse Problems and Imaging, and Mathematics of Computations. Since 2017 he is one of the editors-in-chief of the European Journal of Applied Mathematics. 
\end{IEEEbiography}
%\vspace{-1cm}
\begin{IEEEbiography}[{\includegraphics[width=1in,height=1.25in,clip,keepaspectratio]{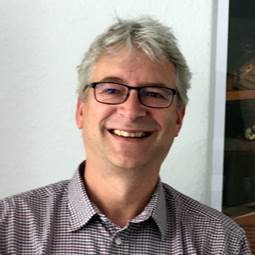}}] {Anko B\"{o}rner} was born in Berlin, Germany. He studied electrical engineering at the Technical University Ilmenau. He joined the German Aerospace Center (DLR) in 1996 for his PhD studies on data processing on-board satellites.
 After receiving the doctoral degree in 1999 he accepted a PostDoc position at Zurich University, Switzerland. In 2000 he came back to Berlin to become a scientific researcher at the German Aerospace Center. Since 2003 he is head of department in different institutes. His research interests are about modelling and simulation of optical systems, computer vision and sensor artificial intelligence. He was involved in several ESA and NASA space missions. 
In 2011 Dr. Börner was awarded with the DLR research grant, which allows him doing a research stay at Auckland, New Zealand. In 2018 he and his team won the “Innovation Award Berlin/ Brandenburg” for developing an optical GPS-free navigation system.
\end{IEEEbiography}
%\vspace{-2cm}
\begin{IEEEbiography}[{\includegraphics[width=1in,height=1.25in,clip,keepaspectratio]{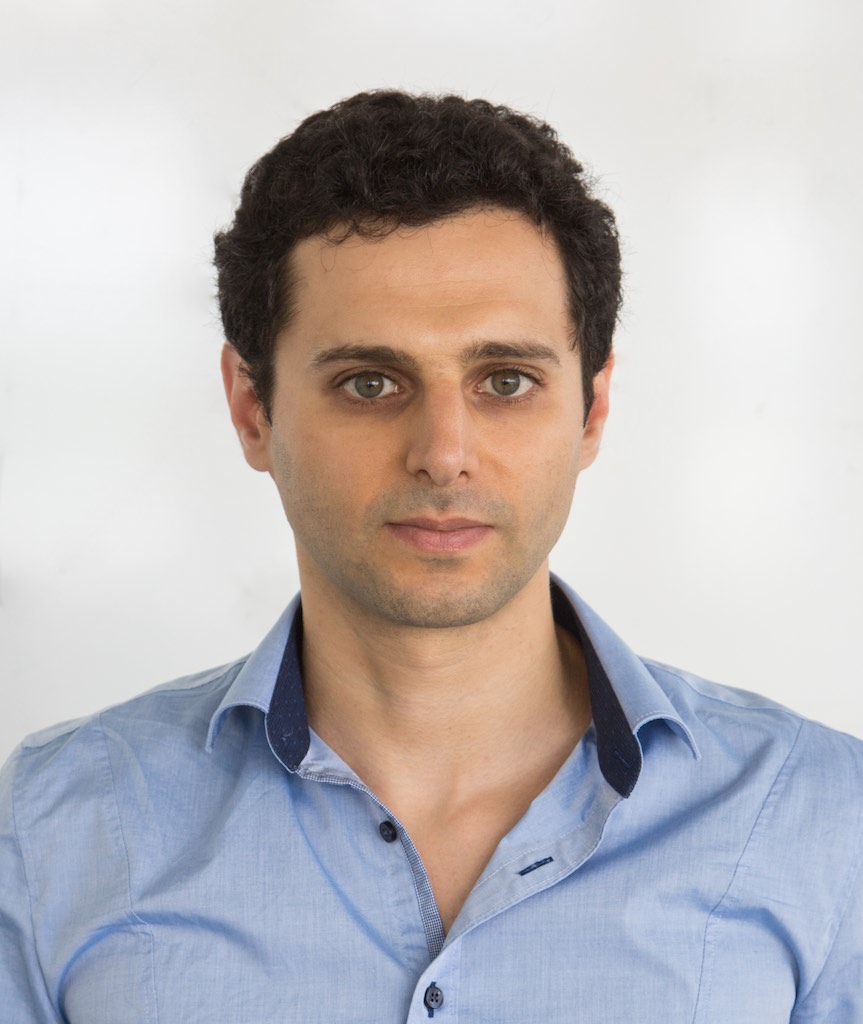}}] {Veniamin I. Morgenshtern} was born in Leningrad, Russia, in 1982. He received the Specialist degree in mathematics and software engineering from Saint-Petersburg State University, Russia in 2004 and the Ph.D. degree in electrical engineering form ETH Zurich, Switzerland, in 2010.
From 2010 to 2012, he was a Postdoctoral Researcher with Electrical Engineering Department at ETH Zurich. From 2012 to 2016, he was a Postdoctoral Researcher with Statistics Department at Stanford University. In 2017 he was a Researcher at Helm.AI, a self-driving car startup. From 2017 to 2018 he was the Chief Scientist at Mentality.AI, an algorithmic consulting company. Since 2018, he has been a Professor in Machine Learning and Signal Processing at Friedrich-Alexander-University Erlangen-Nürnberg (FAU), Germany. His research interests include mathematical signal processing, statistical machine learning, and deep learning. 
Dr. Morgenshtern was a recipient of ETH Zurich Medal for outstanding doctoral thesis in 2011, and of the Fellowship for Advanced Researchers from the Swiss National Science Foundation in 2012. His other awards include the Second Prize in Thomson Reuters Eikon Text Tagging Challenge, a machine learning and natural language processing competition, and the teaching award for the Best Laboratory Course at the Technical Faculty at FAU.
\end{IEEEbiography}

\newpage
\section*{Supplementary Material for Region-of-Interest Prioritised Sampling for Constrained Autonomous Exploration Systems}

\setcounter{figure}{0}
\renewcommand{\thefigure}{\Roman{figure}}
Fig.~\ref{segProc} shows the detection and segmentation step of the acquisition algorithm. Before the detection process the low resolution reconstruction is resized to its macro pixel image dimensions. For example, a~$256 \times 256$ low resolution reconstruction with a macro pixel size of~$4\times4$ is resized to a~$64 \times 64$ image by averaging each~$4 \times 4$ block in the low resolution reconstruction. The resized low resolution reconstruction is the input to the detection and segmentation algorithm with the following steps:
\begin{enumerate}
    \item Define the seed pixel as the brightest pixel in the macro pixel image.
    \item Keep increasing the size of the region by one pixel on each side of the seed pixel until the total contrast of the region is above a user defined threshold. The formed region is one RoI. The definition of contrast used is $\frac{Intensity_{max} - Intensity_{min}}{Intensity_{max} + Intensity_{min}}$, where \textit{Intensity\textsubscript{max}} denotes the maximum intensity within the current region and \textit{Intensity\textsubscript{min}} denotes the minimum  intensity within the current region.
    \item Remove the pixels forming the RoI from the list of all pixels.
    \item Define a new seed pixel as the brightest pixel among the remaining pixels and repeat steps 2 and 3 until the user defined total number of RoIs is reached.
%    \item Merge RoIs if they satisfy a minimum separation distance threshold.
    \item Merge RoIs if they are within the merge radius of each other. The merge radius is a user defined value.
    \item Resize the segmented macro pixel dimension image to the original low resolution reconstruction size, equal to the dimensions of the DMD.
    \item Expand the RoIs to their next dyadic size, subsuming any other RoI that may fall within the extended RoI. For example, a RoI of size $24 \times 24$ will be extended to size $32 \times 32$. This is to allow proper application of dyadic signal transforms in the next steps of the algorithm. Expanding the RoIs also prevents small holes within closely spaced RoIs. 
\end{enumerate}

\begin{figure*}[!t]
\centering
\includegraphics[width=2\columnwidth, keepaspectratio]{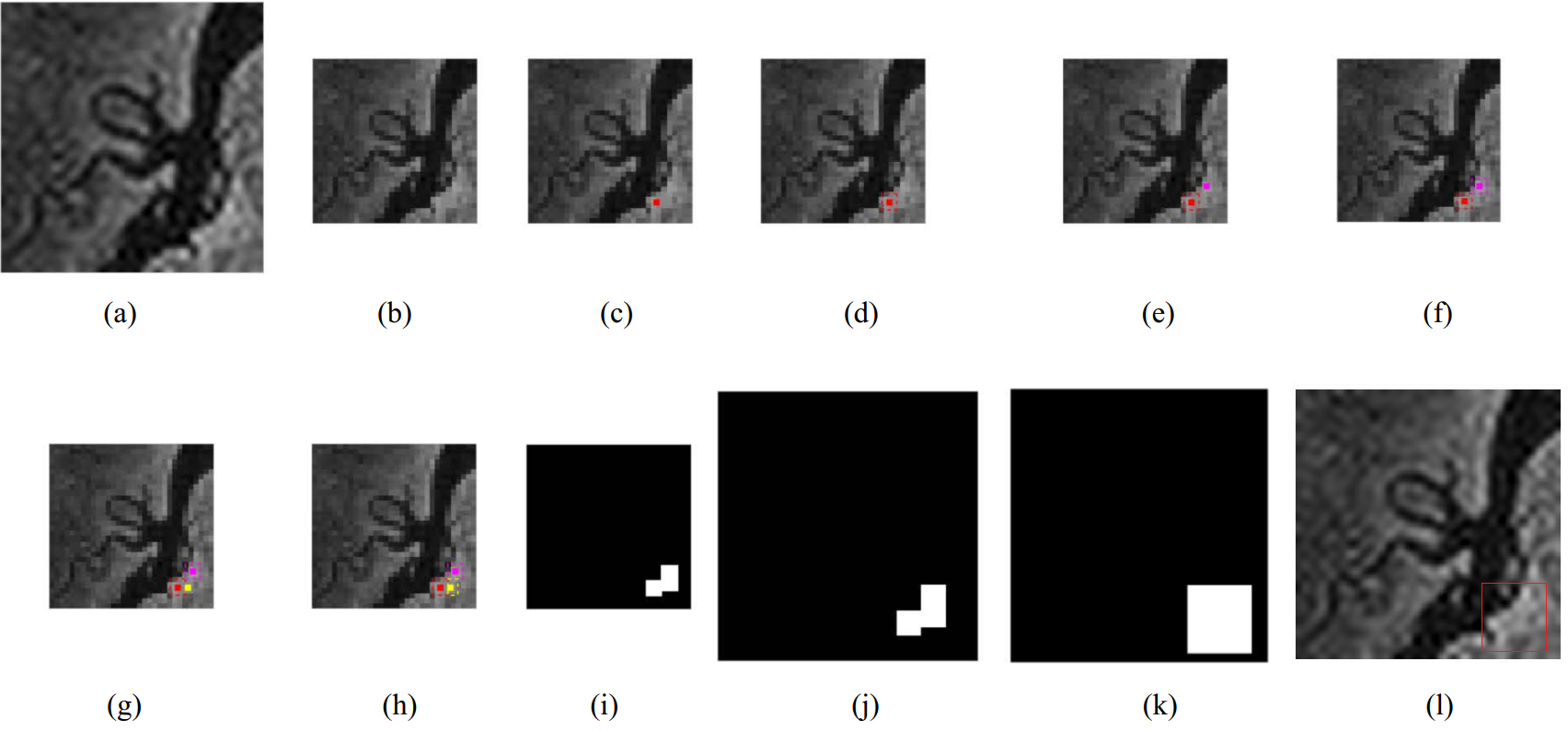}
\caption[Method of RoI detection and segmentation from the coarse reconstruction.]{Method for RoI detection and segmentation from the low resolution reconstruction. (a)~Low resolution reconstruction at original size ($256 \times 256$). (b)~Resized to macro pixel image dimension ($32 \times 32$), the considered macro pixel size is $8 \times 8$. (c)~Seed (brightest) pixel (Red). (d)~Increase the region around the seed pixel by one pixel in all directions. Stop after contrast is above a threshold. (e)~Second seed pixel (Pink). (f)-(h)~Repeat steps 2-4 in the algorithm description. The total number of RoIs in this example is~3. (i)~Segmented Image ($32 \times 32$). (j) Segmented Image resized to original dimensions ($256 \times 256$). (k)~RoIs extended to the next dyadic size. (l)~Representation of RoI on low resolution reconstruction. }
\label{segProc}
\end{figure*}

\begin{figure}[!t]
\centering
\includegraphics[width=\columnwidth, keepaspectratio]{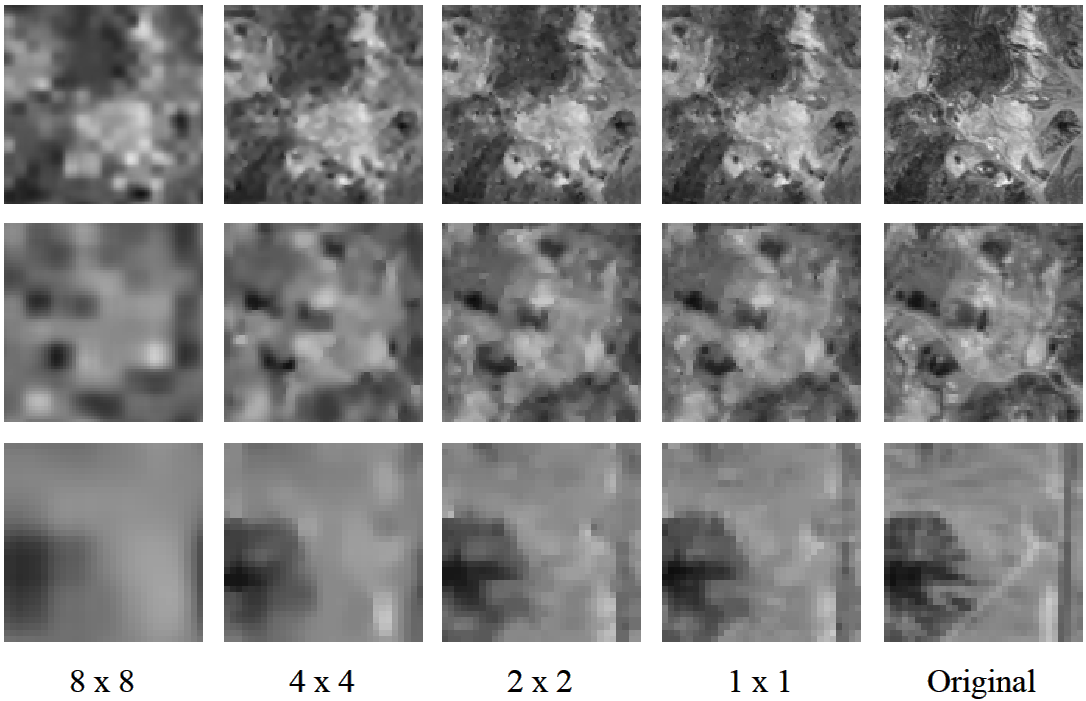}
\caption{Reconstruction of the Cuprite Vis RoIs (Fig.~\ref{figCupVisRI} in the main article) at various macro pixel resolutions. All RoIs are scaled to the same size for visualization. \textit{Top panel}: RoI~1 ~(Red), \textit{middle panel}: RoI~2~(Orange), and \textit{bottom panel}: RoI~3~(yellow).}
\label{figCupVisRoIs}
\end{figure}

Fig.~\ref{figCupVisRoIs} depicts the improvement in the reconstruction of the RoIs with the RPS algorithm for the Cuprite Vis dataset. The error metrics, NMSE and SSIM, are tabulated in the main article in Table~\ref{tableCupVisRI_roiMetric}. 
%\begin{figure}[h]

\begin{figure}[!t]
\centering
\includegraphics[width=\columnwidth, keepaspectratio]{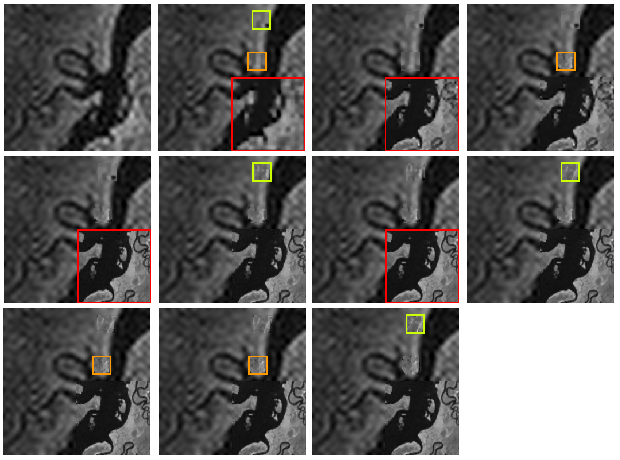}
\caption[Prioritisation of RoIs with RI for Gulf Vis dataset.]{Prioritisation of RoIs with RI for Gulf Vis dataset. The total number of measurements is 8370 (12.8\% of the total number of pixels).}
\label{figGulfVisRI_rois}
\end{figure}

Fig.~\ref{figGulfVisRI_rois} shows the prioritisation of the RoIs with RI for the Gulf Vis dataset and Table~\ref{tableGulfVisRIValues} records the RI values and RoIs selected at each iteration.

%Gulf RI evo table

\begin{table}
\renewcommand\thetable{A}
\caption{Evolution of RI for Gulf Vis dataset. The highlighted RoI indicates the one selected at each iteration. The refined macro pixel resolution of the selected RoI is also tabulated. The total number of measurements is 8370.}
\begin{center}
\newcolumntype{P}[1]{>{\centering\arraybackslash}p{#1}}
\begin{tabularx}{\columnwidth}{|>{\centering}P{0.8cm}|>{\centering}P{1.28cm}|>{\centering}X|>{\centering}X|>{\centering}X|>{\centering\arraybackslash}X|}
\hline
	& \multicolumn{3}{c|}{Refinement Indicator}  & & \\	
	\cline{2-4}
	Iteration No.& \thead{RoI 1\\ (Red)\\ ($128 \times 128$)} & \thead{RoI 2\\ (Orange)\\ ($32 \times 32$)} & \thead{RoI 3\\ (Yellow)\\ ($32 \times 32$)} & Refined Macro Pixel Resolution & Available Measurements\\
	\hline
1 & \cellcolor{green} 508.77 &77.90 &45.54& $4 \times 4$&3686 \\
	\hline
2 & 68.60 &\cellcolor{green}77.90 & 45.54& $4 \times 4$& 2048\\
\hline
3 & \cellcolor{green} 68.60&5.30& 45.54 &$2 \times 2$&1946 \\
\hline
4 & 6.16 & 5.30&\cellcolor{green} 45.54 & $4 \times 4 $ &308\\
\hline
5&\cellcolor{green}6.16 & 5.30  & 5.39&  $1 \times 1$ &206\\
\hline
6 &-&5.30&\cellcolor{green}5.39& $2 \times 2$&206\\
\hline
7 & - & \cellcolor{green}5.30 & 0.56 & $2 \times 2$&104 \\
\hline
8 & -&\cellcolor{green}0.71& 0.56& $1 \times 1$&2\\
\hline
9 & - & - & \cellcolor{green}0.56 &$1 \times 1$&2 \\
\hline
10 & -  & - &- &-&2\\
\hline
\end{tabularx}
\label{tableGulfVisRIValues}
\end{center}
\end{table}

\begin{figure*}[!t]
\centering
\includegraphics[width=2\columnwidth,  keepaspectratio]{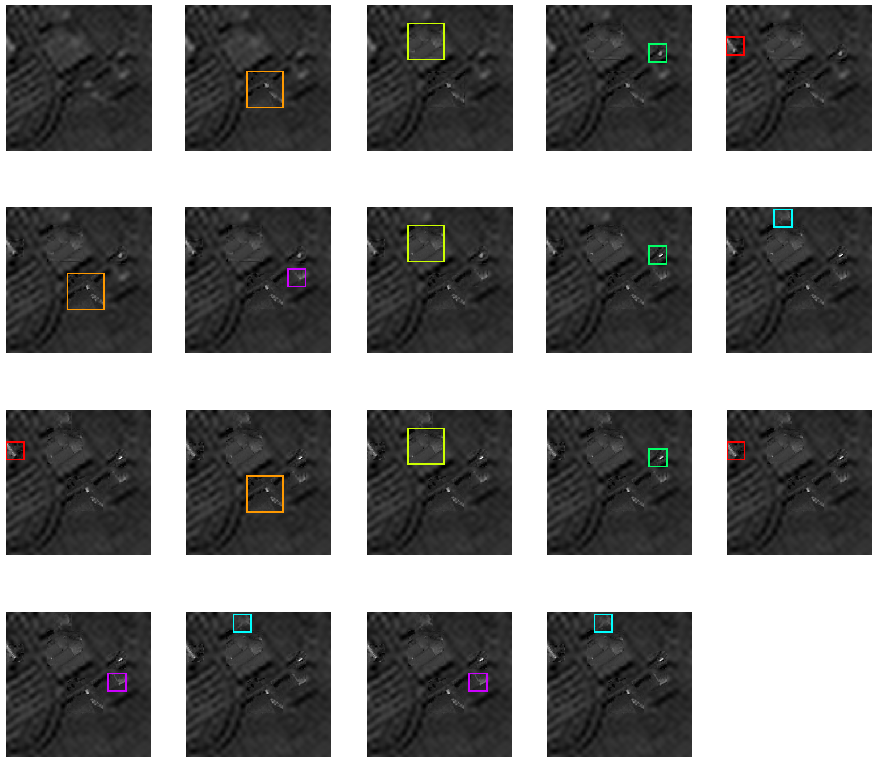}
\caption{Prioritisation of RoIs with RI for Pavia dataset. The total number of measurements is 5910 (9\% of the total number of pixels).}
\label{figPaviaRI_rois}
\end{figure*}

Fig.~\ref{figPaviaRI_rois} shows the prioritisation for the Pavia dataset. The corresponding evolution of the RI is recorded in Table~\ref{tablePaviaRIValues}.

% Pavia

\begin{table*}
\renewcommand\thetable{B}
\caption[Evolution of Refinement Indicator for Pavia dataset.]{Evolution of Refinement Indicator for Pavia dataset in Fig.~\ref{figPaviaRI_rois}. The highlighted RoI indicates the one selected at each iteration. A smaller-sized RoI with more information change is prioritised over a larger RoI with smaller information change across spatial scales.}
\begin{center}
\newcolumntype{P}[1]{>{\centering\arraybackslash}p{#1}}

\begin{tabularx}{\textwidth}{|>{\centering}X|>{\centering}X|>{\centering}X|>{\centering}X|>{\centering}X|>{\centering}X|>{\centering}X|>{\centering}X|>{\centering\arraybackslash}X|}
\hline
	& \multicolumn{6}{c|}{Refinement Indicator}& & \\	
	\cline{2-7}
%	Iteration No.& RoI 1\newline(Red)\newline($32 \times 32$) & RoI 2 (Orange) ($64 \times 64$) & RoI 3  (Yellow) ($64 \times 64$) & RoI 4\newline(Dark Green) ($32 \times 32$)  & RoI 5 (Purple) ($32 \times 32$)  & RoI 6 (Cyan) ($32 \times 32$) & Refined Macro Pixel Resolution &  Available Measurements  \\
%	Iteration No.& RoI 1\newline(Red)\newline($32 \times 32$) & RoI 2 (Orange) ($64 \times 64$) & RoI 3  (Yellow) ($64 \times 64$) & RoI 4\newline(Dark Green) ($32 \times 32$)  & RoI 5 (Purple) ($32 \times 32$)  & RoI 6 \newline(Cyan)\newline($32 \times 32$) & Refined Macro Pixel Resolution &  Available Measurements  \\
Iteration No.& \thead{RoI 1\\(Red)\\($32 \times 32$)} & \thead{RoI 2\\ (Orange)\\ ($64 \times 64$)} &\thead{ RoI 3 \\ (Yellow)\\ ($64 \times 64$)} & \thead{RoI 4 \\(Dark Green)\\ ($32 \times 32$)}  &\thead{RoI 5\\ (Purple)\\ ($32 \times 32$)}  & \thead{RoI 6\\ (Cyan)\\($32 \times 32$)} & Refined Macro Pixel Resolution&  Available Measurements \\
	\hline
1 & 25.69 & \cellcolor{green}33.17 & 14.28&15.69& 13.51&7.04&$4 \times 4$&2458 \\
\hline
2 &\cellcolor{green} 25.69 & 4.19 & 14.28&15.69& 13.51&7.04& $4 \times 4$&2049 \\
\hline
3 &4.05 & 4.19 & 14.28&\cellcolor{green}15.69& 13.51&7.04& $4 \times 4$&1947 \\
\hline
4 &4.05 & 4.19 & \cellcolor{green}14.28&0.96& 13.51&7.04& $4 \times 4$&1845\\
\hline 
5 &4.05 & 4.19 & 2.53&0.96& \cellcolor{green}13.51&7.04& $4 \times 4$&1436 \\
\hline
6 &4.05 & 4.19 & 2.53&0.96& 5.90&\cellcolor{green}7.04& $4 \times 4$&1334\\
\hline
7 &4.05 & 4.19 & 2.53&0.96& \cellcolor{green}5.90&0.54& $2 \times 2$&1232\\
\hline
8 &4.05 & \cellcolor{green}4.19 & 2.53&0.96& 0.42&0.54& $2 \times 2$&1130\\
\hline
9 &\cellcolor{green}4.05 & 0.72 & 2.53&0.96& 0.42&0.54& $2 \times 2$&721\\
\hline
10 &0.81 & 0.72 & \cellcolor{green}2.53&0.96& 0.42&0.54& $2 \times 2$&619\\
\hline
11 &0.81 & 0.72 & 0.66&\cellcolor{green}0.96& 0.42&0.54& $2 \times 2$&210\\
\hline
12 &\cellcolor{green}0.81 & 0.72 & 0.66&0.20& 0.42&0.54& $1 \times 1$&108\\
\hline  
13 & - & \cellcolor{green}0.72 & 0.66&0.20& 0.42&0.54& $1 \times 1$&108\\
\hline
14 & - & - & \cellcolor{green}0.66&0.20& 0.42&0.54& $1 \times 1$&108\\
\hline
15 & - & - & -& 0.20& 0.42&\cellcolor{green}0.54& $2 \times 2$&108\\
\hline
16 & - & - & -& 0.20& \cellcolor{green}0.42&0.12& $1 \times 1$&6\\
\hline
17 & - & - & -& \cellcolor{green}0.20& -&0.12& $1 \times 1$&6\\
\hline
18 & - & - & -& -& -&\cellcolor{green}0.12& $1 \times 1$&6\\
\hline
19 & - & - & -& -& -&-& -&6\\
\hline
\end{tabularx}

\label{tablePaviaRIValues}
\end{center}
\end{table*}
%%%%%%%%%%%%%%%%%%%%%%%%%%%%%%%%%%%%%%

%MSL
Table~\ref{tableMars1RI_roiMetricWalsh} and Table~\ref{tableMars1RI_roiMetricWalsh_limited} record the error metrics for the Mars1 image with the total number of measurements equal to~7552 and~6301, respectively. Table~\ref{tableRIEvolution_Mars2} records the evolution of the RI for Mars2 image whose RoI prioritisation is shown in Fig.~\ref{fig:riEvolution_mars2} in the main article and Table~\ref{tableMars2RI_roiMetricWalsh} shows the corresponding error metrics.

\begin{table}
\renewcommand\thetable{C}
\caption{Error metrics for the RoIs for Mars1 image with Walsh measurements (Fig.~\ref{fig:riEvolution_mars1} in main article) using multi-level sampling in the RPS algorithm.}
\begin{center}
\newcolumntype{P}[1]{>{\centering\arraybackslash}p{#1}}
\begin{tabularx}{\columnwidth}{P{1.1cm}|>{\centering}X|>{\centering}X|>{\centering}X|>{\centering\arraybackslash}X|}
\cline{2-5}
%& \multicolumn{2}{>{\centering\setlength\hsize{2\hsize} }X|}{RoI 1 \newline (Red) \newline($128 \times 128$)}& \multicolumn{2}{>{\centering\setlength\hsize{2\hsize} }X|}{RoI 2 \newline(Orange)\newline ($128 \times 128$)}\\	
& \multicolumn{2}{>{\centering\setlength\hsize{2\hsize} }X|}{\thead{RoI 1\\(Red)\\($128 \times 128$)}}& \multicolumn{2}{>{\centering\setlength\hsize{2\hsize} }X|}{\thead{RoI 2\\(Orange)\\ ($128 \times 128$)}}\\	
	\cline{2-5}
 & NMSE & SSIM & NMSE & SSIM \\
	\hline
\multicolumn{1}{|c|}{Coarse Step}& 0.017 &0.45 &0.011& 0.45  \\
\hline
\multicolumn{1}{|c|}{Refinement Step 1} &0.005 &0.75 &0.005&0.70 \\
\hline
\multicolumn{1}{|c|}{Refinement Step 2}&0.004 & 0.79& 0.004&0.74 \\
\hline
\multicolumn{1}{|c|}{Refinement Step 3}& 0.004& 0.79 & 0.004& 0.75 \\
\hline
\end{tabularx}
\label{tableMars1RI_roiMetricWalsh}
\end{center}
\end{table}

\begin{table}
\renewcommand\thetable{D}
\caption{Error metrics for the RoIs for Mars1 image with Walsh measurements (Fig.~\ref{fig:riEvolution_mars1_limited} in main article) using multi-level sampling in the RPS algorithm with limited measurement budget. The `-' means that the RoI was not resolved.}
\begin{center}
\newcolumntype{P}[1]{>{\centering\arraybackslash}p{#1}}
\begin{tabularx}{\columnwidth}{P{1.1cm}|>{\centering}X|>{\centering}X|>{\centering}X|>{\centering\arraybackslash}X|}
\cline{2-5}
%& \multicolumn{2}{>{\centering\setlength\hsize{2\hsize} }X|}{RoI 1 \newline\hbox{\centering (Red)} \newline \hbox{\centering($128 \times 128$)}}& \multicolumn{2}{>{\centering\setlength\hsize{2\hsize} }X|}{RoI 2\newline (Orange) \newline ($128 \times 128$)}\\	
& \multicolumn{2}{>{\centering\setlength\hsize{2\hsize} }X|}{\thead{RoI 1 \\ (Red) \\($128 \times 128$)}}& \multicolumn{2}{>{\centering\setlength\hsize{2\hsize} }X|}{\thead{RoI 2\\ (Orange)\\ ($128 \times 128$)}}\\
	\cline{2-5}
 & NMSE & SSIM & NMSE & SSIM \\
	\hline
\multicolumn{1}{|c|}{Coarse Step}& 0.017 &0.45 &0.011& 0.45  \\
\hline
\multicolumn{1}{|c|}{Refinement Step 1} &0.005 &0.75 &0.005&0.70 \\
\hline
\multicolumn{1}{|c|}{Refinement Step 2}&0.004 & 0.79& - & - \\
\hline
\multicolumn{1}{|c|}{Refinement Step 3}& 0.004& 0.79 & - & - \\
\hline
\end{tabularx}
\label{tableMars1RI_roiMetricWalsh_limited}
\end{center}
\end{table}

\begin{table}
\renewcommand\thetable{E}
\caption{Evolution of RI for Mars2 image with Walsh measurements using multi-level sampling (Fig.~\ref{fig:riEvolution_mars2} in the main article). The highlighted RoI indicates the one selected at each iteration. The total number of measurements is 5299.}
\begin{center}
\begin{tabularx}{\columnwidth}{|>{\centering}p{0.8cm}|>{\centering}X|>{\centering}X|>{\centering}X|>{\centering\arraybackslash}X|}
\hline
	& \multicolumn{3}{c|}{RI} &\\	
%	\hline
\cline{2-4}
Iteration No.& \thead{RoI 1 \\(Red)\\($128 \times 128$)} & \thead{RoI 2\\ (Orange)\\ ($64 \times 64$)} &\thead{ RoI 3\\ (Yellow)\\ ($32 \times 32$)} & Available Measurements  \\
	\hline
1 & \cellcolor{green} 870.75 & 241.74 & 46.10& 1477 \\
	\hline
2 &  98.10 & \cellcolor{green}241.74& 46.10& 788 \\
\hline
3 & \cellcolor{green}98.10 &  16.48 & 46.10&709 \\
\hline
4 &  22.33 &  16.48  & \cellcolor{green} 46.10 & 147 \\
\hline
5 & \cellcolor{green} 22.33 &  16.48  & 3.21 & 80\\
\hline
6 & - & \cellcolor{green}16.48 & 3.21 & 80 \\
\hline
7 & - & \cellcolor{green} 6.58 & 3.21 & 16 \\
\hline
8 & - & - & \cellcolor{green}3.21 &16 \\
\hline
9 & - & - & \cellcolor{green}2.75 &0\\
\hline
10 & -  & - & - & 0 \\
\hline
\end{tabularx}

\label{tableRIEvolution_Mars2}
\end{center}
\end{table}

\begin{table}
\renewcommand\thetable{F}
\caption{Error metrics for the RoIs for Mars2 image with Walsh measurements (Fig.~\ref{fig:riEvolution_mars2} in the main article) using multi-level sampling in the RPS algorithm with limited measurement budget.}
\begin{center}
\newcolumntype{P}[1]{>{\centering\arraybackslash}p{#1}}
\begin{tabularx}{\columnwidth}{P{1cm}|>{\centering}X|>{\centering}X|>{\centering}X|>{\centering}X|>{\centering}X|>{\centering\arraybackslash}X|}
\cline{2-7}
& \multicolumn{2}{>{\centering\setlength\hsize{2\hsize} }X|}{\thead{RoI 1 \\(Red)\\ ($128 \times 128$)}}& \multicolumn{2}{>{\centering\setlength\hsize{2\hsize} }X|}{\thead{RoI 2\\ (Orange)\\ ($128 \times 128$)}} & \multicolumn{2}{>{\centering\setlength\hsize{2\hsize} }X|}{\thead{RoI 3\\ (Yellow)\\ ($ 32 \ \times \ 32$)}}\\	
	\cline{2-7}
 &  NMSE & SSIM & NMSE & SSIM & NMSE & SSIM \\
	\hline
\multicolumn{1}{|c|}{Coarse Step}& 0.030 &0.44 &0.028& 0.37 & 0.021 & 0.63 \\
\hline
\multicolumn{1}{|c|}{Refinement Step 1} &0.005 &0.84 &0.004&0.88&0.004 & 0.87\\
\hline
\multicolumn{1}{|c|}{Refinement Step 2}&0.004 & 0.87& 0.003 & 0.88& 0.003& 0.91\\
\hline
\multicolumn{1}{|c|}{Refinement Step 3}& 0.004& 0.88 & 0.003 & 0.89 & 0.002 & 0.91\\
\hline
\end{tabularx}

\label{tableMars2RI_roiMetricWalsh}
\end{center}
\end{table}

\begin{figure}
\centering
\includegraphics[ width=\columnwidth, keepaspectratio]{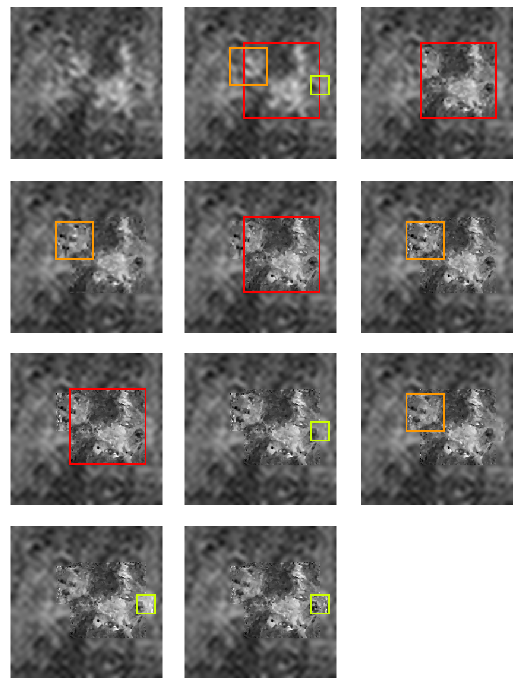}
\caption{Prioritisation of RoIs with RI for Cuprite Vis dataset with random-macro-pixel measurements and additive Gaussian noise. Noise standard deviation is $0.001$.}
\label{figCupVisNoise1}
\end{figure}

\begin{figure}[!t]
\centering
\includegraphics[ width=\columnwidth, keepaspectratio]{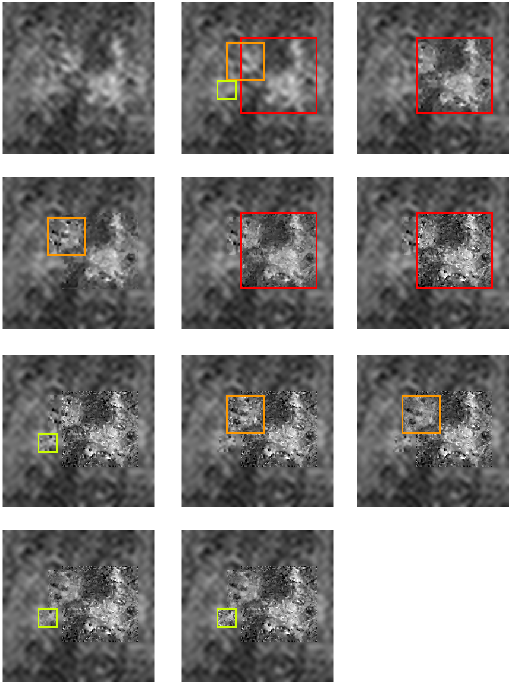}
\caption{Prioritisation of RoIs with RI for Cuprite Vis dataset with random-macro-pixel measurements and additive Gaussian noise. Noise standard deviation is $0.01$.}
\label{figCupVisNoise2}
\end{figure}

\begin{figure}[!t]
\centering
\includegraphics[ width=\columnwidth, keepaspectratio]{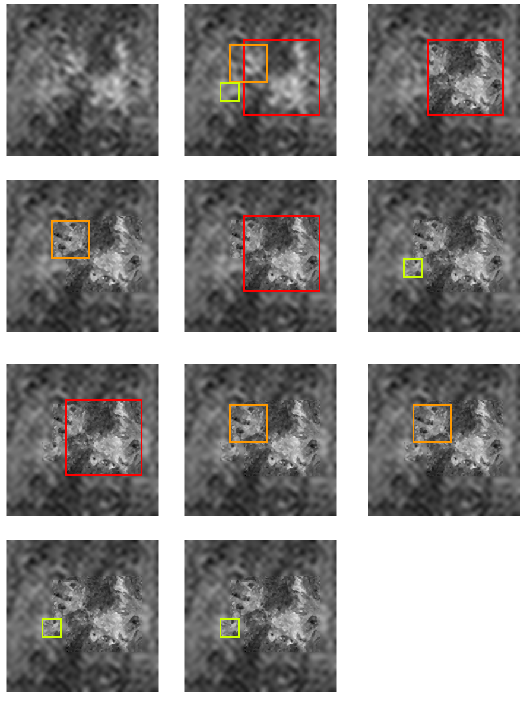}
\caption{Prioritisation of RoIs with RI for Cuprite Vis dataset with Walsh measurements and additive Gaussian noise. Noise standard deviation is $0.1$.}
\label{figCupVisNoise3}
\end{figure}

\begin{figure}[!t]
\centering
\includegraphics[ width=\columnwidth, keepaspectratio]{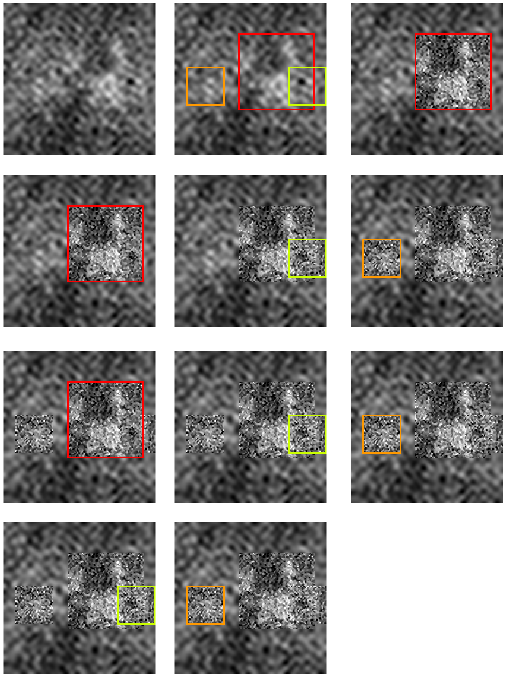}
\caption{Prioritisation of RoIs with RI for Cuprite Vis dataset with Walsh measurements and additive Gaussian noise. Noise standard deviation is $1$.}
\label{figCupVisNoise4}
\end{figure}

Figs.~\ref{figCupVisNoise1},~\ref{figCupVisNoise2},~\ref{figCupVisNoise3}, and~\ref{figCupVisNoise4} show the RPS with noisy measurements. The RPS is applied in the same way as the RI only considers the given data independent of their model. The RoI selections may change with the noise realisations but we expect the first ones to remain more stable. Deviations from idealised solutions are due to model errors rather than our adaptive algorithm. 

\begin{figure}[!t]
\centering
\includegraphics[ width=\columnwidth, height = 8cm,keepaspectratio]{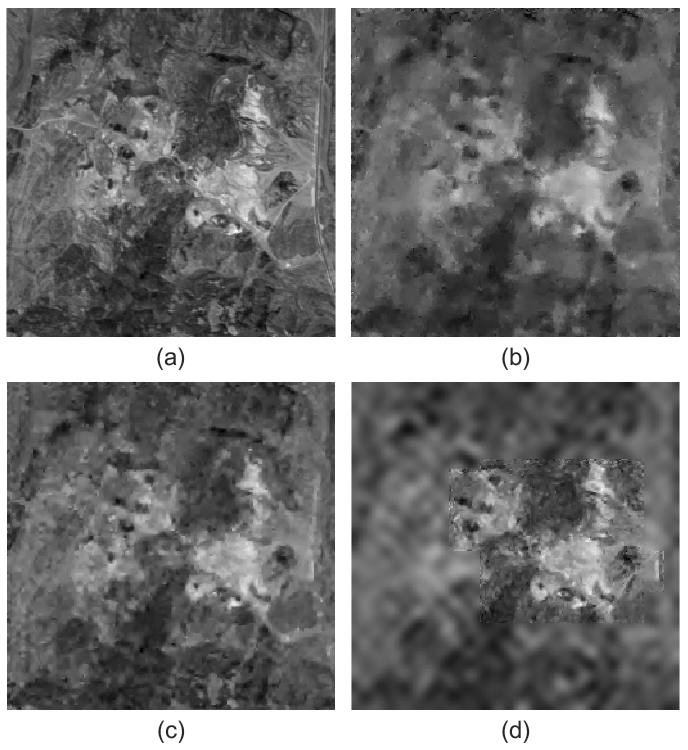}
\caption{Reconstruction comparison of the RPS with classical compressed sensing and multi-level compressed sensing. (a) Original, (b) classical compressed sensing, (c) multi-level compressed sensing, and (d) RPS.}
\label{figCompCupVisCS}
\end{figure}

Fig.~\ref{figCompCupVisCS} shows the comparison between classical compressed sensing, multi-level compressed, and RPS. The error metrics are tabulated in Table~\ref{tableCompCupVisCs} in the main article. 

\end{document}